\documentclass[twoside,twocolumn,9pt]{article}
\usepackage{extsizes}
\usepackage[super,sort&compress,comma]{natbib}
\usepackage[version=3]{mhchem}
\usepackage[left=1.5cm, right=1.5cm, top=1.785cm, bottom=2.0cm]{geometry}
\usepackage{balance}
\usepackage{mathptmx}
\usepackage{sectsty}
\usepackage{graphicx} 
\usepackage{lastpage}
\usepackage[format=plain,justification=justified,singlelinecheck=false,font={stretch=1.125,small,sf},labelfont=bf,labelsep=space]{caption}
\usepackage{float}
\usepackage{fancyhdr}
\usepackage{fnpos}
\usepackage[english]{babel}
\addto{\captionsenglish}{}
\usepackage{array}
\usepackage{droidsans}
\usepackage{charter}
\usepackage[T1]{fontenc}
\usepackage[usenames,dvipsnames]{xcolor}
\usepackage{setspace}
\usepackage[compact]{titlesec}
\usepackage{hyperref}

\usepackage{comment}
\usepackage[normalem]{ulem}

\usepackage[utf8]{inputenc}

\begin{document}

\twocolumn[
\begin{@twocolumnfalse}
\noindent\Large{\textbf{Voltage-driven transition from steady-state fluctuations to phase-transition noise in nanoscale VO$_2$ devices}}\\

\noindent\large{Sebastian Werner Schmid\textit{$^{a,b}$}, Zolt\'an Balogh$^{\ast}$\textit{$^{a,c}$}, Botond S\'anta \textit{$^{a}$}, T\'imea N\'ora Török\textit{$^{a,d}$}, Zsombor Sin\'oros-Szab\'o\textit{$^{a}$}, Gy\"orgy Moln\'ar\textit{$^{d}$}, J\'anos Volk \textit{$^{d}$}, L\'aszl\'o P\'osa \textit{$^{a,d}$}, and Andr\'as Halbritter\textit{$^{a,c}$}}\vspace{0.6cm}

\textit{$^{a}$~Department of Physics, Institute of Physics, Budapest University of Technology and Economics, Műegyetem rkp. 3., H-1111 Budapest, Hungary}\\
\textit{$^{b}$~Experimental Physics V, Center for Electronic Correlations and Magnetism, University of Augsburg, Augsburg 86159, Germany}\\
\textit{$^{c}$~HUN-REN-BME Condensed Matter Research Group, Műegyetem rkp. 3., H-1111 Budapest, Hungary}\\
\textit{$^{d}$~Institute of Technical Physics and Materials Science, HUN-REN Centre for Energy Research, Konkoly-Thege M. \'ut 29-33, H-1121 Budapest, Hungary}\\

$^{\ast}$\textit{Corresponding author: balogh.zoltan@ttk.bme.hu}\vspace{0.6cm} 
\section*{Abstract}
We investigate the electrically driven metal-to-insulator transition (MIT) in nanoscale vanadium dioxide (VO$_2$) Mott memristor through noise spectroscopy and two-dimensional resistor network simulations. Our experiments focus on both the insulating phase as the applied voltage approaches the threshold voltage (set transition) and the metallic phase as the voltage is reduced toward the reset voltage (reset transition). In both regimes, we observe an order of magnitude increase in relative current noise near the transition points. To analyze the origin of this noise enhancement, we use simulations that capture the stochastic dynamics of the phase transition. The simulations indicate that the increased noise stems from amplified phase fluctuations near the percolation threshold, where competing metallic and insulating domains lead to dynamic reconfiguration of the conduction paths. In addition, we show that the precursor current fluctuations observed near the switching threshold are consistent with the threshold voltage variability measured in repeated switching cycles, indicating that the noise sets a lower bound on the achievable variance. These findings offer key insights into the non-equilibrium processes governing phase transitions in nanoscale VO$_2$ devices under electrical stimuli.
\noindent\normalsize{}
\end{@twocolumnfalse} \vspace{0.6cm}]

\section*{Introduction}
Vanadium dioxide (VO$_2$) is a strongly correlated electronic material that shows a reversible insulator-to-metal transition (IMT) near 340~K, accompanied by a structural phase change from a monoclinic insulator to a rutile metal \cite{morin1959oxides, goodenough1971two, zylbersztejn1975metal}. This transition can be triggered using various external stimuli such as electric fields \cite{Psa2021,ko2008,delValle2019}, optical pulses \cite{li2022,Kim2025}, or strain \cite{Wu2006,Cao2009,Tian2018,Fang2022}, making VO$_2$ a promising candidate for applications in neuromorphic computing, memory devices, and tunable electronics \cite{Son2012,Jung2021,Gao2022,Carapezzi2022,Haddad2022,Maher2024,Schmid2024}.
Among many other applications, VO$_2$ devices have been successfully used to build circuits that reproduce various neural spiking patterns \cite{Yi2018} and to assemble oscillatory neural networks that can solve optimization problems, such as map coloring or maximum cut.\cite{Dutta2021,Mohseni2022,Maher2024,molnar2026neural,pollner2026vo2}

The rich neurodynamic properties of VO$_2$ circuits are accompanied by a complex physical behavior of the individual devices.\cite{Kalcheim2020}
Previous studies have shown that the electronically triggered insulator-to-metal transition (the so-called set transition) involves the nucleation and growth of metallic domains within the insulating matrix \cite{Qazilbash2007}. 
As the applied voltage increases toward a critical threshold, these domains percolate, leading to an abrupt change in conductance when a metallic path is formed \cite{Sohn2015}. The reverse process occurs during the so-called reset transition, where the system reverts to the insulating phase as the voltage is reduced. These transitions rely on the rich interplay of thermal phenomena,\cite{Psa2023} nonlinear electronic effects,\cite{Kalcheim2020,Gopalakrishnan2009} and subtle dynamical processes that occur near the switching thresholds, especially those involving temporal and spatial fluctuations of the coexisting phases.\cite{Stinson2018,Tiwari2026}

Noise spectroscopy offers a sensitive tool to go beyond the investigation of \emph{average} resistive switching characteristics, and to capture the fluctuations in the active volume.\cite{Balogh2021,balogh2023configuration,santa2019universal,santa2021noise,posa2021noise,nyary2025benchmarking} On the one hand, this is very important for characterization, as an excessive 1/f-type noise may fundamentally limit the device performance. On the other hand, noise spectroscopy is a useful diagnostic tool that aids the understanding of the key physical processes during the device operation. Despite the extensive interest in VO$_2$ as a resistive switching material, only a few studies have focused on characterizing its noise properties, mainly focusing on the investigation of single crystals and bulk thin film samples \cite{Topalian2015, Gunes2024, andreev1980low, almeida2000, baidakova1997structural, basantani2012enhanced, jelks1975response, velichko2003deterministic}.
In this work, we use nanoscale VO$_2$ devices, where the operation is focused to an ultra-small, $30~$nm-scale spot,\cite{Psa2023,Schmid2024,pollner2026vo2} and we apply a specially designed noise measurement setup that not only records the noise spectrum of selected device states, but also maps the variation of noise characteristics throughout the entire resistance switching cycle. This allows us to track how the noise precursors the switching process, i.e., how the noise amplitude increases significantly as the switching thresholds are approached. The comparison of the measured noise characteristics to electrothermal resistor network simulations uncovers the dominance of intrinsic resistance fluctuations further away from the switching voltages, which defines a base noise level with weak voltage dependence for both the metallic and insulating states. As the switching thresholds are approached, however, phase fluctuations are becoming dominant, giving rise to an order of magnitude noise increase.
Comparing the noise levels of our nanoscale VO$_2$ devices with those of other resistive switching systems, we find that the low-resistance states (LRS) exhibit noise characteristics similar to filamentary memristive systems operating in the diffusive conduction regime. In contrast, the high-resistance states (HRS) show a significantly lower relative noise level than that typically observed in other memristive systems in a similar resistance range, reflecting bulk-like transport in VO$_2$ versus fluctuation-enhanced conduction in filamentary systems with reduced active volumes. Furthermore, the increased noise observed in the vicinity of the switching threshold is consistent with the switching-threshold variance, indicating that device noise plays a key role in determining the threshold variability.

\section*{Results and discussion}
\begin{figure}[h!]
    \centering
    \includegraphics[width=\columnwidth]{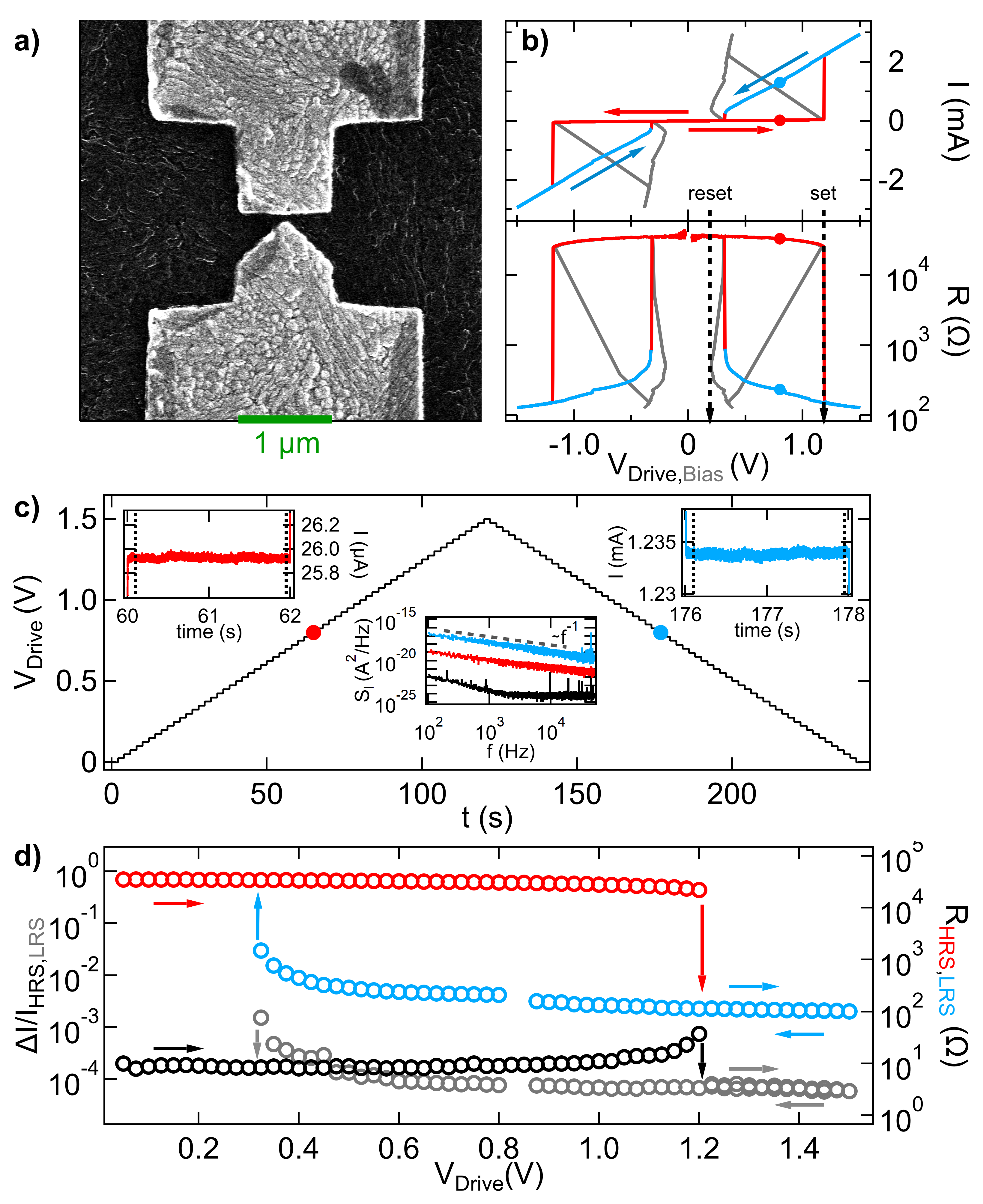}
    \caption{\textbf{Current-voltage and noise characterization of the VO$_2$ memristors.} 
    a) SEM image of the fabricated platinum contacts (gray) forming an $\approx40$~nm wide gap on the top of the thin VO$_2$ film (black). The green scalebar demonstrates $1~\mu$m distance. b) Current and resistance curves measured with a 380$~\Omega$ serial resistance. The arrows indicate the ramping direction of $V_\mathrm{Drive}$ for the insulating (red) and metallic (blue) state. The gray line shows the $V_\mathrm{Bias}$ dependence. The red and blue dot mark the current and resistance for the HRS and LRS at $V_\mathrm{Drive}=0.8$~V respectively. c) Voltage signal applied during the noise spectroscopy. Each plateau has a length of $t=2~$s and a voltage step of $\Delta V=25$~mV. The insets on the upper left and right hand corners show the $I(t)$ traces for the HRS (red) and the LRS (blue) at $V_\mathrm{Drive}=0.8~$V as indicated by the dots in b). The dotted black lines signify the start and end time to evaluate the power spectral densities shown in the inset at the center of the figure for the HRS (red) and LRS (blue), where the gray line marks a $f^{-1}$ slope. The black curve is a reference PSD at 0~V.
    d) Resistance (red, blue) and relative current noise (black, gray) as a function of $V_\mathrm{Drive}$ with a serial resistance of $R_\mathrm{s}=440~\Omega$. For $825~\mathrm{mV}\le V_\mathrm{Drive}\le850~\mathrm{mV}$ a small resistance jump occurs and as a result the PSDs are not dominated by the flicker noise. Because this was repeatedly observable, we concluded that this must originate from a local inhomogeneity in the LRS for that voltage region. Therefore, we exclude these data points for the noise evaluation as the other spectra did not show this phenomena.}
    \label{fig1}
\end{figure}

In our experiments, we used thin VO$_2$ films created by thermal oxidation of a vanadium layer and evaporated Pt contacts on top which form an asymmetric nanogap shown in Fig.~\ref{fig1}a. With this sample geometry electrode separations of $30-40~$nm can be achieved, and the V-shaped electrode on one side focuses the resistive switching to a localized nanoscale spot. Further details of the sample preparation are available in the Methods Section and in our previous publication. \cite{Psa2021}

Applying a triangular voltage signal with a sufficiently high amplitude results in unipolar and volatile resistive switching (RS) of the device as displayed in Fig.~\ref{fig1}b. Starting at $V_\mathrm{Drive}=0$~V, the system remains in the insulating high resistance state (HRS) until a threshold voltage, the so-called set voltage is reached ($V_\mathrm{Set}\approx 1.2$~V in Fig~\ref{fig1}b), where the active VO$_2$ volume switches to a metallic low resistance state (LRS). This switching leads to a distinct decrease of the device resistance, which, together with the serial resistance $R_\mathrm{s}=380~\Omega$ causes a redistribution of the $V_\mathrm{Bias}=V_\mathrm{Drive}-I\cdot R_\mathrm{s}$ bias voltage on the device, yielding a negative differential resistance (NDR) during the transition. Upon voltage reduction, the NDR is observed a second time, when the drive voltage reaches the so-called reset voltage ($V_\mathrm{Reset}\approx 0.2$~V in Fig.~\ref{fig1}b), where the MIT switches the device back to its HRS and the share of $V_\mathrm{Drive}$ increases at the device. The top panel of Fig.~\ref{fig1}b shows the $I(V)$ curve of this resistive switching cycle both as a function of $V_\mathrm{Drive}$ (red and blue curves) and $V_\mathrm{Bias}$ (gray curve). In the former case, the red (blue) segments indicate the HRS (LRS) of the device. The arrows indicate the direction of the voltage sweep. The bottom panel in Fig.~\ref{fig1}b shows the resistance vs.\ voltage dependence $(R(V)=I(V)/V)$ for the same measurement with the same marking. Note, that the the set and reset voltages can be determined from the gray curves, as shown by the vertical dashed lines in Fig.~\ref{fig1}b.   

Compared to simple $I(V)$ measurements, where the current can be measured during the continuous ramping of the voltage, the noise spectroscopy demands constant voltage levels. This is necessary because otherwise variations of the current would be dominated from voltage changes instead of actual fluctuations. This problem can be eliminated by changing the voltage in a stepwise manner and evaluating the noise spectra along the steps, omitting the transient ranges. Fig.~\ref{fig1}c shows such a stepwise voltage signal applied for the noise measurements. The top insets illustrate the measured current responses to the voltage steps indicated by the red and blue dots in the main panel. From these current segments, we numerically calculate the power spectral density (PSD) of the current noise ($S_I(f)$) in the regions between the dashed lines. The resulting $S_I(f)$ noise spectra are shown in the lower panel by the red and blue curves, while the black curve shows the zero-voltage background noise level. From these raw spectra, we obtain the relative current noise $\Delta I/I$ as follows: (i) we subtract the base noise level measured at $0~$V, which is dominated by the thermal noise of the device and the amplifier; (ii) we calculate the current fluctuation by numerically integrating the spectrum for the selected bandwidth between 100 Hz and 50 kHz, $\Delta I=\sqrt{\int S_\text{I}\text{d}f}$, and (iii) we normalize the $\Delta I$ current fluctuation with the average current value $I$ measured along the given voltage plateau. With this evaluation, we obtain a good measure of the overall fluctuation, noting that $\Delta I/I$ should be voltage independent for steady state fluctuations, while a deviation from this constant behavior is indicative of nonlinear noise features,\cite{Balogh2021,posa2021noise} or the excitation of non-steady state fluctuations.\cite{nyary2025benchmarking} This approach, however, discards the spectral information contained in the raw $S_I(f)$ curves. According to our experience, the spectra consistently show an $S_I(f)\sim 1/f$ flicker noise behavior (see the dashed guide-to-the-eye line in the bottom inset of Fig.~\ref{fig1}b), which is characteristic to a larger ensemble of fluctuators with a broad distribution of fluctuation time constants.\cite{Balogh2021}

It is noted, that the stepwise voltage variation illustrated in Fig.~\ref{fig1}c enables the evaluation of the noise characteristics throughout the entire resistance switching cycle. In addition to this so-called full-cycle noise measurement, the $I(V)$ curve can also be evaluated from the same data (average current values along the voltage plateaus), i.e., the noise and $I(V)$ characteristics can be compared for the same voltage ramp. An example of such a full-cycle noise measurement is illustrated in Fig.~\ref{fig1}d, where the measurements were performed on the same device states as those shown by the continuous $I(V)$ measurement in Fig.~\ref{fig1}b. The red and blue curves reproduce the $R(V)$ measurements shown in the bottom panel of Fig.~\ref{fig1}b using the step-wise measurement method, while the black and gray curves show the voltage dependence of the relative noise $\Delta I /I$ in the  HRS and LRS, respectively. The direction of the voltage sweep is indicated by the arrows along the curves.  

It can be stated, that the the $\Delta I/I$ relative noise levels are mostly voltage independent further away from the transition voltages, i.e. in the HRS well below $V_\mathrm{Set}$, and in the LRS well above $V_\mathrm{Reset}$. However, as the transition voltages are approached, i.e. as the voltage is increased towards $V_\mathrm{Set}$ in the HRS or when is is decreased towards $V_\mathrm{Reset}$ in the LRS, a significant, order-of-magnitude relative noise increase is observed. Accordingly, the increase of the relative noise level acts as a precursor of the transition. This precursor effect is especially prominent in the HRS, where the relative resistance change up to the set voltage is minor, $\left(R\big|_{V=0}-R\big|_{V=V_\mathrm{Set}}\right)/R\big|_{V=0}\approx 0.37$, while the relative change of $\Delta I/I$ is an order of magnitude higher, $\left(\frac{\Delta I}{I}\big|_{V=V_\mathrm{Set}}- \frac{\Delta I}{I}\big|_{V=0}\right)/\frac{\Delta I}{I}\big|_{V=0}\approx 3.24$. In the LRS, the relative change of $\Delta I/I$ is even higher, $(\frac{\Delta I}{I}\big|_{V=V_\mathrm{Reset}}- \frac{\Delta I}{I}\big|_{V\gg V_\mathrm{Reset}})/\frac{\Delta I}{I}\big|_{V\gg V_\mathrm{Reset}}\approx 25$, however, in the LRS the resistance also changes significantly, $\left(R\big|_{V=V_\mathrm{Reset}}- R\big|_{V\gg V_\mathrm{Reset}}\right)/R\big|_{V\gg V_\mathrm{Reset}}\approx 14$, where the $V\gg V_\mathrm{Reset}$ reference values are read at $V_\mathrm{Drive}=1.5~$V. In the following we analyze the origin of these nonlinear noise features separately for the HRS (next section) and for the LRS (section after the next section).

\subsection*{High Resistance State}
\begin{figure}[h!]
    \centering
    \includegraphics[width=\columnwidth]{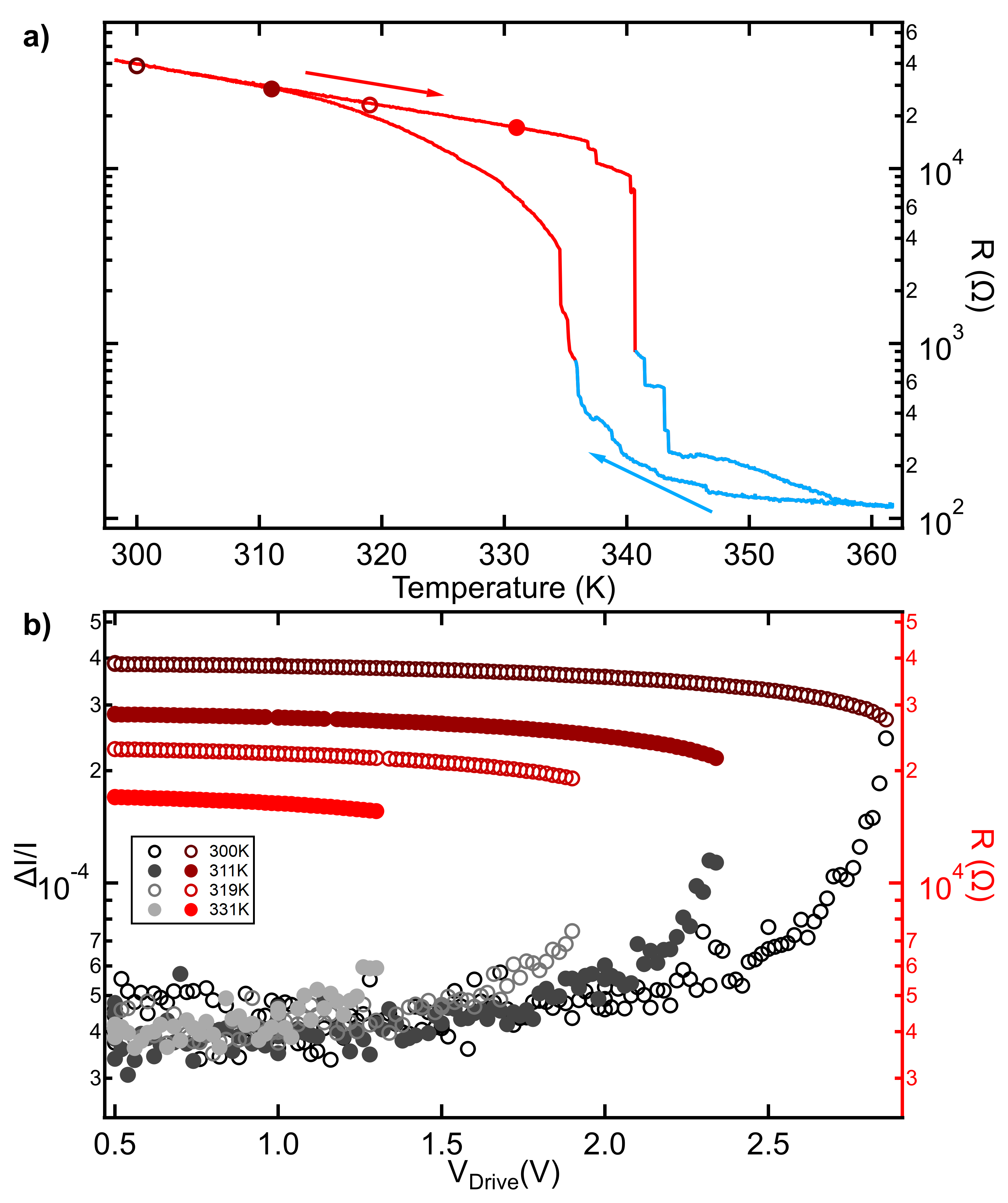}
    \caption{\textbf{The influence of the device temperature on the resistance and current noise.} a) The temperature dependent resistance of VO$_2$ probed with a low readout voltage $V_\mathrm{Drive}=10$~mV. The arrows indicate the ramping direction of the temperature. b) Resistance (red) and relative current noise $\frac{\Delta I}{I}$ (black) of the insulating state as functions of $V_\mathrm{Drive}$ at different temperatures. The low voltage resistance for the different temperatures is indicated by the red dots in a).}
    \label{fig2}
\end{figure}

First, we examine the noise in the high resistance state. To separate voltage-dependent phenomena from purely temperature-dependent features, we first record the temperature dependence of the resistance ($R(T)$) at low readout voltage ($V_\mathrm{Drive}=10$~mV). The hysteretic $R(T)$ curve measured on a device similar to the one demonstrated in Figs.~\ref{fig1}a,b is plotted in Fig.~\ref{fig2}a. The thermally driven transition at $T_\mathrm{c}\approx338$~K is not as sharp as the voltage-driven transition observed in Figure~\ref{fig1}b, where the formation of a metallic phase leads to an increase in current, resulting in positive feedback on the dissipated Joule heat. Instead, it shows stepwise resistance changes that cover a range of $\approx15$~K. This may be related to the distribution of the transition temperature $T_\mathrm{IMT}$ within the layer, the stochastic nature of the transition process, or even additional effects on the energy of the system, such as strain or interface energy.\cite{Wu2006,cahn1958free,Rocco2022}  It can also be seen that prior to the phase transition the resistance of the sample decreases linearly with increasing temperature on the logarithmic plot. Along this part of the $R(T)$ curve we choose dedicated base temperature values (see the circles), where we perform noise measurements. The voltage dependent resistance (red) and relative noise (gray) for different base temperatures are shown in Figure~\ref{fig2}b. All these measurements are performed up to the highest possible voltage values before the set transition is achieved. The stepwise voltage sweep is performed with $\Delta V_\mathrm{Drive}=20~$mV voltage resolution, i.e., we approach the set voltage with this resolution. Obviously, at higher base temperature the set transition is achieved at lower voltage.

Concerning the voltage dependence of the $R(V)=V_\mathrm{Bias}/I$ resistance (red circles in Fig.~\ref{fig2}b), a modest resistance decrease is observed further away from $V_\mathrm{Set}$, while the nonlinearity is more pronounced in the close vicinity of the transition, in accordance with our previous studies.\cite{Psa2023} The offset of the curves at the different base temperatures is consistent with the temperature dependence of the low-voltage resistance shown by the circles in Fig.~\ref{fig2}a.

In the low voltage region ($V_\mathrm{Drive}<1~$V) the relative noise data (gray curves in Fig.~\ref{fig2}b) are mostly voltage independent. This is characteristic to steady-state fluctuations of the device conductance $S_ G$, which yields current fluctuations of $S_I=S_ G\cdot V$. The latter relation yields voltage-independent relative current fluctuations, which equal the steady-state relative conductance or resistance fluctuations, $\Delta I/I=\Delta G/G=\Delta R/R$.\cite{Balogh2021} As a sharp contrast to the factor of two variation of the base resistance at different base temperatures (see the offset of the red curves), the relative noise values do not depend on the base temperature in the $V_\mathrm{Drive}<1~$V region within the error of the measurement. This indicates that the intrinsic resistance fluctuation of the VO$_2$ device has the same temperature dependence as the resistance itself, so that $\Delta R(T)\propto R(T)$,  i.e.
\begin{equation}
\frac{\Delta R(T)}{R(T)}=\mathrm{const}. 
\label{eq:const}    
\end{equation}

The relative noise curves for different base temperatures show a similar common trend at higher voltages as well. However, we see a significant deviation from this common trend near the set voltage on each curve, where the noise increases significantly. At the lowest base temperature, this increase spans almost an order of magnitude, which is again a significantly greater relative change than what we observe in the temperature dependence of resistance, similarly to Fig.~\ref{fig2}b. These highly elevated noise levels near $V_\mathrm{Set}$ seem to be a \emph{precursor} of the insulator-to-metal transition. In the following, we use a two-dimensional resistor network model to examine the possibility of describing this precursor noise as phase transition noise resulting from the simultaneous presence of two phases.

\begin{figure}[h!]
    \centering
    \includegraphics[width=\columnwidth]{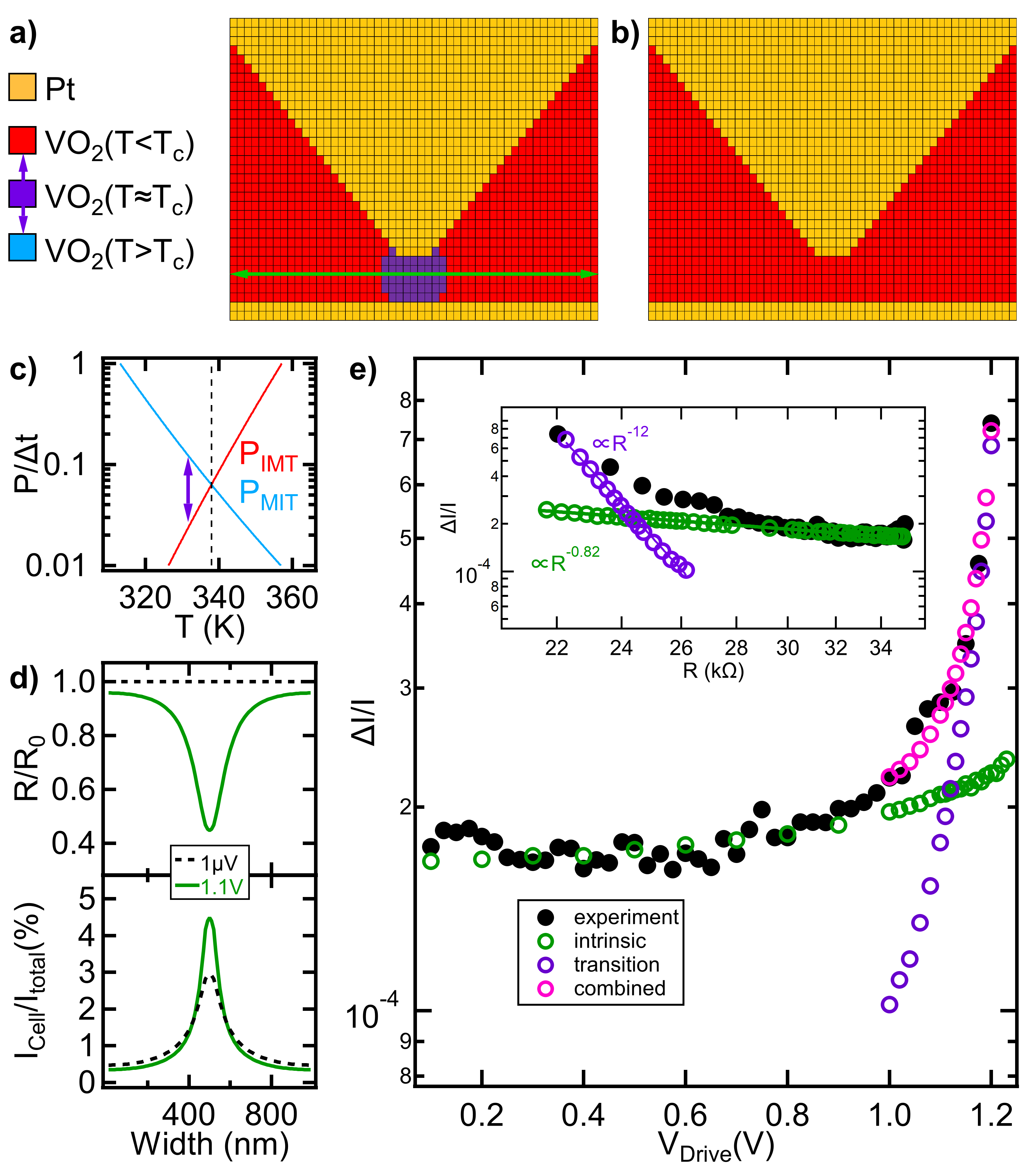}
    \caption{\textbf{Simulating the noise signal of the HRS in the 2-dimensional resistor network.}\\ 
    a-b) Schematic structure of the sample geometry used in the two-dimensional resistor network simulations (not to scale). Yellow squares indicate the metallic contacts, red ones are insulating VO$_2$ and purple squares can be either insulating or metallic. While a) indicates the local cell temperature at high voltages close to $V_\mathrm{Set}$, b) shows the case for low voltages. c) Temperature dependence of the switching probability per timestep of the IMT (red) and MIT (blue) used to simulate the transition. d) Local resistance and relative current of the cells indicated by the green line in a) during the simulations. The resistance depends on the local electric field and temperature which both contribute to the simulated resistance at elevated voltages (green) compared to the starting resistance $R_0$ at room temperature and zero electric field. This resistance variation leads to a focusing of the current close to the gap at high voltages (green) compared to low voltages (black). e) Voltage dependence of the relative current noise from experiments (black) and simulations of the transition noise (purple) as well as the intrinsic noise (green), which combined (magenta) follows the trend of the experimental data of the HRS. The inset shows the resistance scaling of the relative noise.}
    \label{fig3}
\end{figure}

The two-dimensional resistor network simulations were carried out for a 1~$\mu \mathrm{m}^2$ area of the sample geometry schematically shown in Figure~\ref{fig3}b using $5\cdot5\cdot40~\mathrm{nm}^3$ vanadium dioxide cells. Here, the electrodes are shown in yellow and the vanadium-dioxide cells in red. For each VO$_2$ cell, temperature ($T$) and electric field ($E$) dependent resistance ($R_{M/I}(T,E)$) was assumed depending on its state (metallic or insulating). The local electric field and temperature of the individual cells were obtained by solving the heat and circuit equations for the resistor network, whose parameters were determined partly from literature and measurement data, and partly by adjusting them so that the simulated voltage dependence of the resistance fits well with the measured $R(V)$ curve. For further details, see the Methods or Sections~1.1 and 2.1 of the ESI. 

To model resistive switching in the simulations, transition probabilities $P_\mathrm{IMT}$ and $P_\mathrm{MIT}$ between two phases were introduced with finite values below and above the transition temperature $T_\mathrm{c}$, respectively, such that $P_\mathrm{IMT} = P_\mathrm{MIT}$ at $T = T_\mathrm{c}$ (see Fig.~\ref{fig3}c). This results in an insulating (metallic) phase of the cells for $T \ll T_\mathrm{c}$ ($T \gg T_\mathrm{c}$), while the state can switch between the two near $T \approx T_\mathrm{c}$. The transition probabilities were calculated using a Landau free-energy approach with linearly temperature-dependent energy gaps.\cite{Rocco2022} The temperature range of finite transition probabilities was chosen to be consistent with the hysteresis width shown in Fig.~\ref{fig2}a (see Sections~1.2 and 2.2 of the ESI for details on the calculation and choice of probabilities).

During the simulation, the voltage applied on the resistor network causes Joule heating to dissipate across the cells, resulting in a location-dependent change in the temperature. At a given voltage, after the thermalization, this sets an equilibrium temperature and resistance for each cell. The corresponding evolution of the cells in HRS is illustrated in Fig.~\ref{fig3}a,b. At sufficiently low voltages, the entire VO$_2$ region is in the insulating state (red squares in Fig.~\ref{fig3}b). As the voltage is increased, the local temperature increases and because of the V-shaped electrode geometry, the temperature change is largest in the central region. At sufficiently high voltage ($V \approx V_\mathrm{Set}$) the local temperature of individual cells approaches $T \approx T_\mathrm{c}$. Consequently, due to the finite $P_\mathrm{IMT}$ and $P_\mathrm{MIT}$, stochastic phase transitions of the cells start to occur (illustrated by purple squares in Fig.~\ref{fig3}a).  At even higher voltages ($V > V_\mathrm{Set}$), the temperature of some cells reaches the $T \gg T_\mathrm{c}$ regime, stabilizing their states in the metallic phase (light blue squares in Fig.~\ref{fig5}a,b). As the voltage further increases, the number of cells in the metallic state (i.e., the metallic volume) also increases, and only the side of this region is fluctuating between the two states (purple region in Fig.~\ref{fig5}a). Then, as the voltage decreases, the volume of the metallic region begins to shrink. First, it reduces to the size of the gap (Fig.~\ref{fig5}b), and finally, when  $V < V_\mathrm{Reset}$, the system returns to the HRS.

The observed stochastic nature of the phase transition introduces a fluctuation into the system, referred to as \emph{phase transition noise}. In addition to this transition-based fluctuation, the \emph{intrinsic resistance fluctuations} of the cells are also taken into account by adding a random fluctuation term to the resistance:
\begin{equation}
R(T,E,x) = R(T,E)\cdot(1+x\mathcal{N}),
\label{Eq:R_noise}
\end{equation}
where $\mathcal{N}$ is the standard normal distribution and $x$ refers to the fluctuation strength. Note that this choice yields a temperature and voltage-independent relative resistance fluctuation (called \emph{intrinsic noise}) for each cell in accordance with Eq.~\ref{eq:const}.

Fig.~\ref{fig3}e shows the experimental relative current fluctuation for the HRS as a function of voltage together with simulation results for three cases: (i) solely intrinsic noise (green), (ii) solely transition noise (purple), and (iii) combined intrinsic and transition noise (magenta). The comparison of the simulated and experimental noise shows that, by combining intrinsic and transition noise, the simulation is able to reproduce the experimentally observed voltage dependence, i.e., constant relative noise at low voltage, a slight increase at moderate voltage, and finally a steep increase in the proximity of $V_\mathrm{Set} $ (see Section~2.2 of the ESI for details of the fitting of the simulation results to the experiment).

By investigating the separated simulated cases (green and purple in Fig.~\ref{fig3}e), the microscopic evolution of noise can be better understood. At low voltage, the intrinsic noise follows the expected constant relative noise level in the linear I(V) regime,\cite{Balogh2021} whereas at slightly higher voltages, the overall relative fluctuation becomes voltage-dependent despite voltage-independent cell fluctuations. This behavior can be explained geometrically. Fig.~\ref{fig3}d shows the cell resistances and the cell contributions to the overall current across the horizontal green line in Fig.~\ref{fig3}a. The black dashed lines show these quantities at a very low readout voltage ($1~\mu$V), where all the cells have the same resistance (top panel), and the central cells give a larger current contribution than the more remote ones (bottom panel). An increased voltage ($1.1~$V; see green lines) yields self-heating in the central region, accompanied by a decrease in the resistance of the central cells (top panel). For this reason, a larger proportion of the current flows through the central cells (bottom panel). Since the contribution of fewer cells dominates the current, the device’s effective active volume decreases, and fluctuations are less averaged out, leading to a slight increase in relative noise. This is closely related to Hooge’s law,\cite{Kogan1996} which states that the relative noise level scales inversely with the dominant device volume, assuming uniformly distributed fluctuations. 

The \emph{quasi-geometric} noise described above explains well the voltage dependence of the relative noise in the moderate voltage region, where the transition of the cells does not yet play a role, but it deviates from the measured data near the threshold voltage. However, the precursor noise (i.e., the steep increase near $V_\mathrm{Set}$) observed in the experiments can be explained well by transition noise (purple markers). In this range, the local temperature of the gap region approaches $T_\mathrm{c}\approx338~$K, therefore the increased $P_\mathrm{IMT}$ and decreased $P_\mathrm{MIT}$ allow the cells to switch between insulating and metallic states, leading to fluctuations of the device resistance, which we observe as noise in the current. At $T\approx T_\mathrm{IMT}$, the transition probabilities become similar, shifting the dominant phase from insulating to metallic cells, which leads to positive feedback on the current and Joule heat. As this sharply increases the local temperature at the gap, the number of fluctuating cells, and thus the relative noise, also increases, until the remaining insulating cells transition to the LRS and form a conducting path between the two metal electrodes, which finalizes the IMT. It is important to mention that although the simulated intrinsic and transition noise reproduce the voltage dependence of the experimental noise at low voltages and close to $V_\mathrm{Set}$, respectively, they fail in the moderate voltage region. However, their combination (pink curve in Fig.~\ref{fig3}e) already gives results consistent with the experimental data (black circles) without further parameter tuning. 

Since the resistance of the system also changes with increasing voltage, it is worth examining the relative fluctuation as a function of resistance, as shown in the inset of Fig.~\ref{fig3}e for the experimental data (black) and for the simulations in the cases of intrinsic (green) and transition (purple) noise. From this, the power-law scaling of $\Delta I/I$ as a function of $R$ can be extracted, yielding a relatively low scaling exponent ($\approx -0.82$) for intrinsic noise and a significantly higher exponent  ($\approx -12$) for transition noise. In the former case, the slight slope ($0.5-1.5$, depending on the dimensionality and distribution of the fluctuators) is expected from a simple geometrical model,\cite{Balogh2021,Fehervari2024,nyary2025benchmarking} in which changes in the active volume amplify the relative noise. In a simple ohmic system, however, the resistance would increase as the active volume shrinks. In our VO$_2$ system, the shrinking of the active volume is caused by temperature changes and has a quasi-geometric origin (i.e., the entire volume contributes to conduction, but the weight of the middle region increases), both leading to a decrease in resistance rather than an increase, resulting in a negative exponent. In contrast, the high exponent in the latter case cannot be captured by such a model but instead can be related to the formation and decay of percolation paths.\cite{Ahn2021}

\subsection*{Low Resistance State}

Next, we examine the noise characteristics of LRS. A fundamental difference between the LRS and the HRS is that the LRS can only be achieved at a finite voltage, i.e., it is not possible to probe its transport and noise characteristics starting from zero bias. In addition, unlike the case of the HRS, in the LRS not only the noise but also the resistance shows an order of magnitude change as we approach the switching threshold (see the blue curve in Fig.~\ref{fig1}d). This raises the question of whether this nonlinearity is due to intrinsic transport nonlinearity, or whether increasing the voltage causes an increasingly larger volume to switch to a conductive state, thereby changing the resistance as a function of voltage. From the point of view of simulations, it is also important to know whether voltage-dependent nonlinearity must be taken into account in the resistance of individual cells in the metallic state. 

\begin{figure}[h!]
    \centering
    \includegraphics[width=\columnwidth]{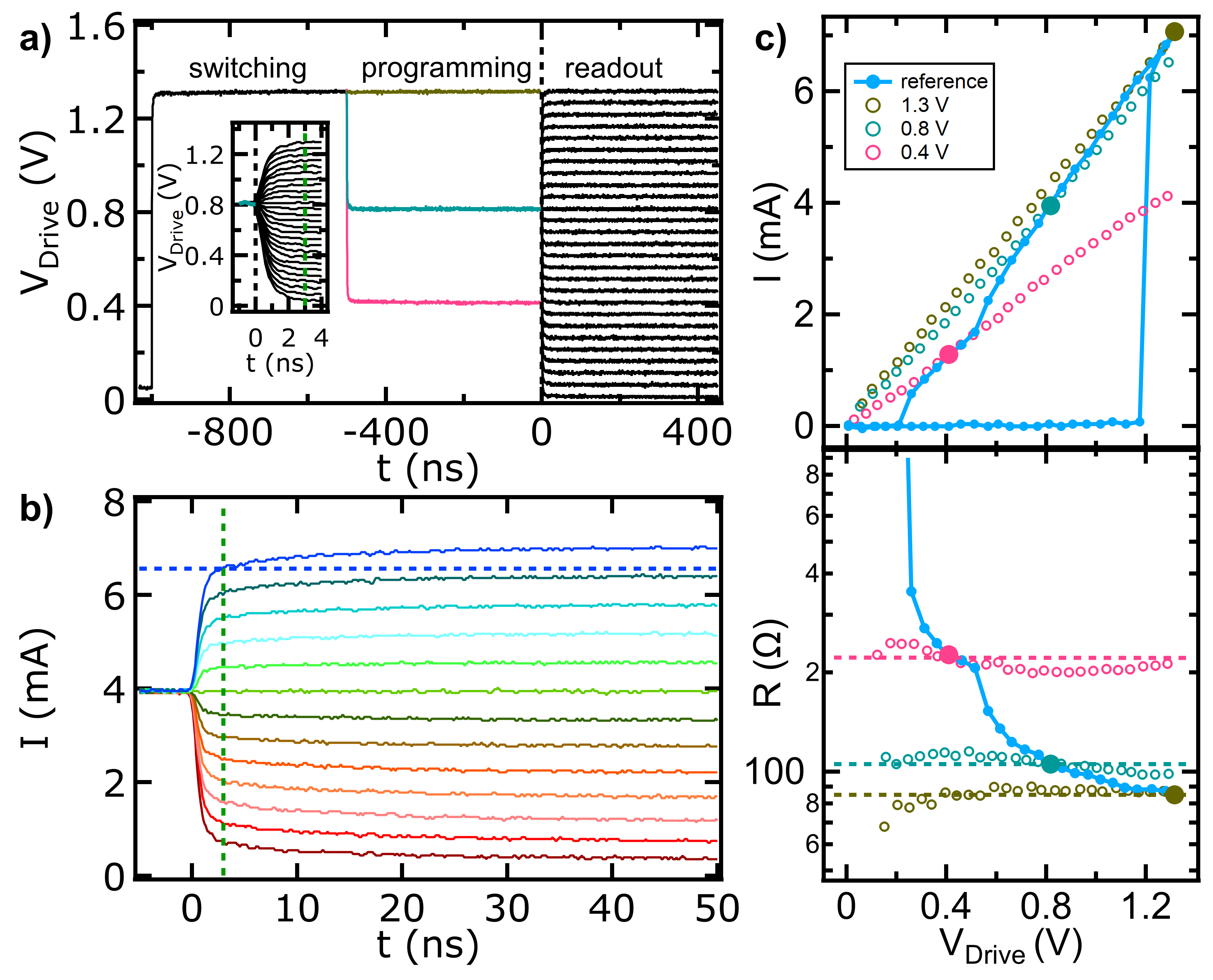}
    \caption{\textbf{Investigation of the voltage  dependence of resistance in LRS.} a) Voltage pulse sequence applied on our VO$_2$ nanostructure to investigate the voltage dependence of the resistance in the LRS. The sequence contains an initial switching pulse, driving the system from HRS into LRS, a set pulse, to settle the sample into a resistive state, and readout pulses to measure the resistive response of the set state under varying voltage inputs. The inset magnifies the transition from the 800~mV set pulse to the readout pulses. The dotted black line marks the starting time of the voltage change and the green dotted line indicates the time $t=3$~ns of the readout for figure c). b) Time evolution of the current during the readout pulses after the 800~mV set pulse from a). The vertical green line marks the same time $t=3$~ns as in the inset of a), while the horizontal blue line indicates the readout current for the highest voltage (1.3~V) at 3~ns. Notably, the current still increases after the $t=3$~ns mark, while the drive voltage is already constant. c) Voltage dependence of the current and resistance of the LRS after different set pulses at the readout time indicated in a) and b). The reference curve is taken at the end of the readout pulses where the resistances have relaxed into their final state. Notably, the current, at same $V_\mathrm{Drive}$, is higher compared to the DC-I(V) in Figure~\ref{fig1}b due to the lack of an additional serial resistance besides the $50~\Omega$ in- and output resistance of the oscilloscope and waveform generator respectively.}
    \label{fig4}
\end{figure}

To clarify these issues,  
we first probe the nonlinear features of the LRS by fast pulsed measurements (see the scheme in Fig.~\ref{fig4}a).\cite{Schmid2024} An initial switching pulse with an amplitude higher than $V_\mathrm{Set}$ is used to initialize the formation of the LRS, while the following programming pulse drives the system to a constant resistance. Afterward, read-out pulses of varying amplitudes are applied. The programming pulse is long enough to achieve a settled resistance, but afterwards the device state is read only $3~$ns after the leading edge of the readout pulses. This $3~$ns is expected to be short in relation to the thermal time constant, meaning that there is no significant shrinkage or expansion in the active volume, but rather the intrinsic transport nonlinearity of the of the actual device state is examined. Note that the various readout levels are applied in consecutive measurements so that after each readout pulse a long waiting time ($100~\mu\mathrm{s}$) is applied at 50~mV to ensure the complete relaxation of the device to its original HRS. The inset in Fig.~\ref{fig4}a magnifies the leading edge of the readout pulses, which shows that the 3~ns readout time already exceeds the rise time of the pulses. 

In the example measurement shown in Fig.~\ref{fig4} we have applied 3 different programming levels, i.e. setting three different device states (see the pink, turquoise and olive programming levels in Fig.~\ref{fig4}a.  Fig.~\ref{fig4}b illustrates the current response to the readout pulses after the $V_\mathrm{Drive}=800~$mV (turquoise) programming pulse. Note, that the driving pulses are applied from an arbitrary waveform generator with $50~\Omega$ output resistance, while the current response is measured by recording the $V_\mathrm{Trans}$ transmitted pulse on a $50~\Omega$ input impedance oscilloscope ($I=V_\mathrm{Trans}/50~\Omega$, see Methods for more details). The current values are read at $3~$ns delay, as shown by the vertical green dashed line. The such obtained $I(V_\mathrm{Drive})$ and $R(V_\mathrm{Drive})$ curves are plotted by the pink, turquoise and olive symbols for the the three different programmed states in Fig.~\ref{fig4}c. These $I(V)$ curves are highly linear demonstrating the lack of any significant intrinsic transport nonlinearity. As a comparison, the light blue curves in Fig.~\ref{fig4}c demonstrate the $I(V_\mathrm{Drive})$ and $R(V_\mathrm{Drive})$ traces such that the current is read at the end of the readout pulses ($2~\mu\mathrm{s}$ delay) instead of the $3~$ns delay. This longer time is enough for the shrinkage or expansion of the active volume according to the actual readout voltage level and the related Joule heating. The such obtained $I(V_\mathrm{Drive})$ and $R(V_\mathrm{Drive})$ already resemble the standard D.C. measurements (see Fig.~\ref{fig1}b,d), showing an order of magnitude nonlinearity in the $R(V_\mathrm{Drive})$ dependence. 

\begin{figure}[h!]
    \centering
    \includegraphics[width=\columnwidth]{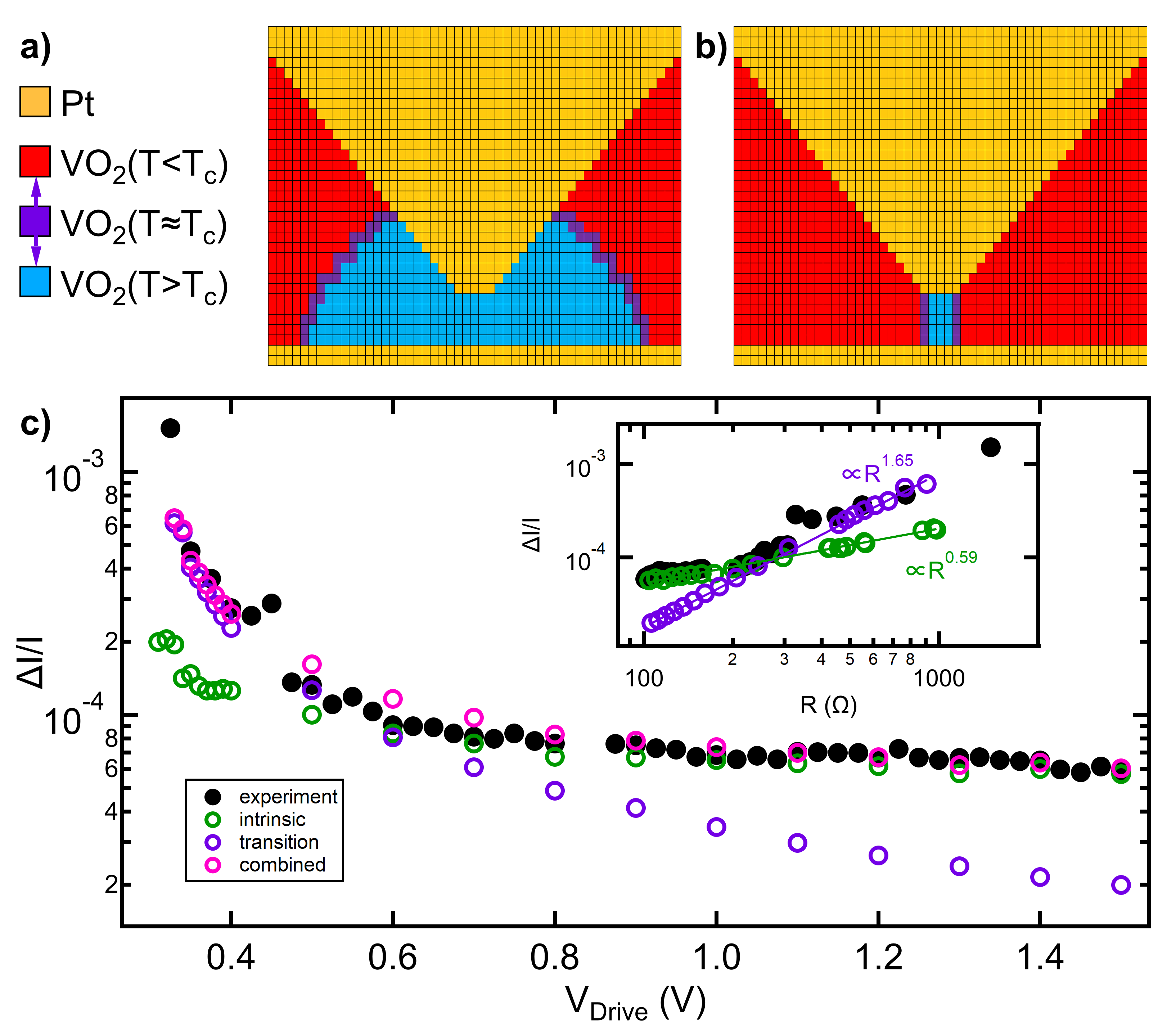}
    \caption{\textbf{Simulating the noise signal of the LRS in the 2-dimensional resistor network.} a-b) schematic structure of the sample geometry used in the two-dimensional resistor network simulations (not to scale) for high voltages (a) and low voltages (b). Yellow squares indicate the metallic contacts, red is insulating VO$_2$, blue is metallic VO$_2$ and purple squares can be either insulating or metallic. c) Voltage dependence of the relative current noise from experiments (black) and simulations of the transition noise (purple) as well as the intrinsic noise (green), which combined (magenta) follows the trend of the experimental data of the LRS. The inset shows the resistance scaling of $\frac{\Delta I}{I}$.}
    \label{fig5}
\end{figure}

Utilizing the conclusions of the previous fast pulsed measurements, we carry out the two-dimensional resistor network simulations using voltage-independent cell resistances for the LRS with only a small temperature dependence, originating from electron correlations in the LRS~\cite{qazilbash2008electrodynamics,belozerov2011evidence}, which we implemented as a pseudo activation gap $E_{\mathrm{g,LRS}}=80$~meV (see Section~2.1 of the ESI). Similarly to the HRS, in the LRS we also consider an intrinsic noise, which is a constant relative resistance fluctuation (see Eq.~\ref{eq:const}), while for cells close to the transition temperature, the phase transition noise is also incorporated. In the LRS, the resistance is determined by the size of the metallic region, as the active (metallic) volume varies with the applied bias, rather than by temperature variations or nonlinear transport effects. Consequently, the resistance dependence of the noise is primarily set by the geometry of the metallic phase. Figure~\ref{fig5}a (b) shows the schematic formation of the LRS under high (low) external voltages, where the dissipated Joule heat drives a large (small) volume of VO$_2$ into the metallic phase, leading to a lower (higher) resistance. Again, the red (light blue) cells are in the insulating (metallic) state, while the purple cells fluctuate between the two states. Compared to the HRS, where phase fluctuations appear only for $V\approx V_\mathrm{Set}$, these fluctuations now appear at an interface region between the metallic and insulating volume for any voltage that maintains the LRS (see the purple cells in Fig.~\ref{fig5}a,b).

The results of the simulations in the LRS are compared with the experimental noise data (black circles) in Fig.~\ref{fig5}c. Similar to the HRS, three cases were investigated. The green (purple) curve show the voltage dependence of the relative noise for the case where only intrinsic (phase transition) noise is taken into account, while the pink curve corresponds to the results obtained when both noise types are combined. Again, this latter simulated noise curve is consistent with the experimental data, and further conclusions can be drawn by analyzing the separate cases. The intrinsic noise dominates at voltages well above $V_\mathrm{Reset}$, where the metallic volume is so large that the phase fluctuations are driven further away from the central region and thus have a low impact on the total noise. In this voltage regime, the scaling of the relative noise with the device resistance is weak, exhibiting an exponent of $\approx 0.59$ (see the green curve in the inset). This exponent is consistent with a simple geometrical model of uniformly distributed noise sources in a planar (quasi two-dimensional) device configuration, which predicts a scaling of $\Delta I/I \sim R^{0.5}$, where changes in the active region are determined by variations in its width (see Section~4 of the ESI). In addition, as the voltage is decreased towards $V_\mathrm{Reset}$, the metallic volume shrinks and the fluctuating interface migrates into the gap region. As a consequence, the phase transition noise becomes the dominant contribution, which already shows a stronger power law dependence on the device resistance with an exponent of $\approx 1.65$ (see the purple curve in the inset). This is in good agreement with a good-conductor bad-conductor transition in a two dimensional system, where the bad conductor percolates in the good conductor phase, leading to an exponent of $\approx1.5$\cite{Topalian2015, kiss1993new, hui1986noise}. 

\subsection*{Interpretation of the results}
The simulation results showed that, sufficiently far from the threshold voltages, the observed noise level can be well described by the intrinsic resistance fluctuations of the cells. Comparing this intrinsic noise level of our VO$_2$ devices with the noise levels typically reported for other memristors,\cite{santa2019universal,santa2021noise,Fehervari2024,nyary2025benchmarking,Lombardo2024,Lee2011,Puglisi2016} it can be concluded that, in the $100$--$200\,\Omega$ resistance range of the LRS, the intrinsic relative noise exhibits a similar level ($\sim 10^{-4}$) to that observed in filamentary systems containing relatively thick (few nm wide) filaments. Furthermore, in the transition-noise regime (up to $1~\mathrm{k}\Omega$ resistance and $\sim 10^{-3}$ relative noise), a resistance scaling $\Delta I/I \sim R^{1.65}$ is observed, similar to that reported for the diffusive conduction limit of filamentary systems. \cite{santa2019universal,santa2021noise} However, the relative noise level ($\approx 10^{-4}$) observed in the HRS is significantly lower than the relative noise levels ($10^{-3}$ to $10^{-1}$) in other memristive systems in a similar $20$--$30~$k$\Omega$ resistance range. This regime generally belongs to the broken-filament state of filamentary systems,\cite{posa2021noise,Balogh2021} where the active region is small and the transport becomes barrier-like, leading to enhanced fluctuations. In contrast, transport in VO$_2$ remains bulk-like over a large active region, resulting in a lower noise level.

In synapse-type memristors, steady-state noise is one of the limiting factors for the achievable conductance resolution.\cite{Rao2023,nyary2025benchmarking} However, in neuron-type memristors, such as VO$_2$ devices, the stability of the switching threshold values is the more relevant technological parameter, therefore, it is important to examine how this stability relates to the observed noise. Based on measurement series involving multiple switching events (see Section~5 of the ESI), the $V_\mathrm{Set}$ ($V_\mathrm{Reset}$) values exhibit relative variations of $1$~\% ($2~\%$), while endurance measurements performed on the VO$_2$ oscillator circuit show an approximately $3~\%$ shift in the oscillation frequency.\cite{pollner2026vo2} These values are consistent with the transition-related precursor noise observed in the vicinity of the threshold voltage, characterized by $\approx 2~\%$ ($\approx 4~\%$) relative peak-to-peak fluctuations, corresponding to  relative noise level of $8\cdot 10^{-4}$ ($1\cdot 10^{-3}$). This indicates that, in nanoscale VO$_2$ devices where other processes (e.g., structural rearrangements) do not influence the threshold values, transition noise sets a lower bound on the achievable threshold variance.

\section*{Conclusions}
In conclusion, we have investigated the insulator-to-metal transition (IMT)-based volatile resistive switching in VO$_2$ thin-film nanodevices using nonlinear noise spectroscopy, a technique particularly sensitive to small-scale systems. We observed a well-defined voltage dependence of the relative current noise, including a pronounced increase near the set and reset voltages and a weaker voltage dependence far from these transition points. In addition to the noise measurements, conventional DC methods were used to determine the temperature and voltage dependence of the HRS, while high-frequency measurements in the LRS enabled the separation of voltage-~and~volume-dependent contributions to the sample resistance.

To gain further insight into the microscopic processes underlying the IMT, the experimental results were complemented by two-dimensional resistor-network simulations based on a Landau free-energy framework. The simulations reproduce the experimentally observed noise scaling and capture the distinct contributions of intrinsic and transition-related fluctuations. Furthermore, the simulations provide a microscopic interpretation of the precursor noise observed near the transition voltages and reveal the different scaling behaviors associated with quasi-geometric fluctuations and the formation of transient percolation paths.

Together, the experiments and simulations provide new insight into the nondeterministic nature of the IMT in VO$_2$ and establish noise spectroscopy as a sensitive probe for investigating the microscopic processes underlying the phase transition. Our analysis further shows that the intrinsic noise of nanoscale VO$_2$ devices remains comparatively low in the HRS despite the large device resistance, reflecting the extended, bulk-like character of transport in contrast to filamentary systems and indicating that intrinsic noise is not expected to limit device operation. Moreover, the magnitude of the precursor transition noise was found to be consistent with the cycle-to-cycle variability of the switching thresholds, suggesting that transition noise imposes a lower limit on the achievable threshold variance in nanoscale VO$_2$ devices. These results underscore the importance of noise measurements in the investigation of correlated electron systems, complementing conventional measurement techniques and highlighting the microscopic complexity of fluctuations accompanying first-order transitions.

\section*{Methods/Experimental Section}
\subsection*{Device Fabrication}
The thin film VO$_2$ was fabricated by thermal oxidation of metallic vanadium layer \cite{Psa2021}. The exposure of $400\,^\circ$C during a time of $4.5$ hours led to the formation of a 40~nm thick VO$_2$ film on top of a 180~nm thick V$_2$O$_5$ bottom layer. The metallic contacts, consisting of 10~nm Ti and 50~nm Pt, were
patterned by electron-beam lithography and deposited by
electron-beam evaporation.
\subsection*{DC Characterization}
The DC data of Figures~\ref{fig1}b/~\ref{fig2}a, were measured using a triangular/constant voltage signal applied by a NI USB-6363. This voltage signal was applied on the sample with a serial Resistance of 330~$\Omega$ and the output current was amplified by a Femto DLPCA-200, which adds an additional 50~$\Omega$ to the measurement path. The amplified current is then read out by the same NI USB-6363. The Bias voltage $V_\mathrm{Bias}$ at the sample is then calculated as $V_\mathrm{Bias}=V_\mathrm{Drive}-I\cdot380\Omega$.
\subsection*{RF Characterization}
For the investigation of the LRS we applied the voltage signal shown in Figure~\ref{fig4}a with a Zurich Instrument HDAWG Arbitrary Waveform Generator. There we use $V_\mathrm{Drive}=2\cdot V_\mathrm{In}$, where $V_\mathrm{In}$ is the incoming voltage wave at the sample. The incoming and transmitted voltage signal $V_\mathrm{Trans}$ was measured using a Rhode\&Schwarz RTO1014 digital storage oscilloscope. The current through the sample is the given by $I=V_\mathrm{Trans}\cdot Z_0$, where $Z_0=50~\Omega$ for our $50~\Omega$ impedance matched setup. The sample impedance $Z$ is then calculated as
\begin{equation}
Z=\frac{2Z_0\cdot V_\mathrm{In}}{V_\mathrm{Trans}}-2Z_0
\end{equation}
where negligible capacitive and inductive contributions lead to the sample resistance $R=Z$.
\subsection*{Noise Spectroscopy}
The voltage signal for the noise spectroscopy shown in Figure~\ref{fig1}c was applied by an Agilent 33220A. A low-pass filter was used to separate the instrumental noise from the DC-signal, which is then applied on the sample. The combined serial resistance in the setup was $440~\Omega$ which is comparable to the experiments for DC characterization. The resulting current signal was amplified with a Femto DLPCA-200 and read using a NI PXI-5922 digital oscilloscope, which uses an anti-aliasing filter to subtract noise contributions from frequencies above our measurement bandwidth. The power spectral density $S_\mathrm{I}$ is then calculated by the fast Fourier transformation of the current fluctuation $I(t)-\langle I\rangle$. We also measure a reference PSD $S_\mathrm{I,0}$ at $V_\mathrm{Drive}=0$~V, which consists of the thermal noise of the sample as well as contributions of the amplifier and oscilloscope. Finally, we calculate the current noise as $\Delta I=\sqrt{\int (S_\mathrm{I}-S_\mathrm{I,0})\cdot \mathrm{d}f}$.
\subsection*{Two-Dimensional Resistor Network Simulations}
The resistance of the $i$-th cell, which is either in HRS or LRS, depends on the local temperature ($T_{i}$) and electric field ($E_{i}$), and was calculated as \begin{equation}\label{Eq:R_model}
    R_{i}(T_{i},E_{i})=R_i(T_0,0) \cdot \exp\left( \frac{E_{\mathrm{g},i}}{2 \, \mathrm{k}_\mathrm{B} T_{i}}- \frac{E_{\mathrm{g},i}}{2 \, \mathrm{k}_\mathrm{B} T_0} - \frac{E_{i}}{E_{\mathrm{c},i}} \right)~,
\end{equation}
where $R(T_0,0)$ is the resistance at the base temperature $T_0$ in the absence of an electric field, $k_\mathrm{B}$ is the Boltzmann constant, $E_\mathrm{g}$ is the activation gap energy and $E_\mathrm{c}$ denotes the critical electric field.\cite{Psa2023}
The $R(T_0,0)$, $E_\mathrm{g}$, $E_\mathrm{c}$ model parameters depend on the state of the cells (metallic or insulating) and were determined by the fitting of experimental $R(V)$ and $R(T)$ curves (for further details and the values of the parameters see Section~2.1 and Table~1 of the ESI).

Applying a voltage on the resistor network leads to Joule heat $P_{i}=I_{i}^2R_{i}(T_{i},E_{i})$ in each cell. The local temperature change $\Delta T_{i}$ of a cell during one simulation step $\Delta t$ is then described as
\begin{equation}
    \frac{\Delta T_i}{\Delta t}=\frac{P_i}{C_i}-\frac{G_s}{C_i}(T_i-T_0)-\frac{1}{C_i}\sum_{j}^{NN}G_{ij}(T_i-T_j)~~,
\end{equation}
where $C$ is the heat capacity, $G_s$ and $G_{ij}$ are the thermal conductance to the substrate and the nearest neighbors with $T_0$  and $T_j$ being their respective temperature. Note that the parameters were determined for the insulating and metallic states from the literature and by fitting the simulation to the experiment (for further details and the values see Section~2.1 and Table~1 of the ESI).
For the phase transition probabilities $P$ we used
\begin{equation}
    P_\mathrm{IMT}=\exp\left(\frac{-E_\mathrm{IMT}(T)}{\mathrm{k}_\mathrm{B}T}\right)~~~;~~~P_\mathrm{MIT}=\exp\left(\frac{-E_\mathrm{MIT}(T)}{\mathrm{k}_\mathrm{B}T}\right)~~~,
\end{equation}
with the temperature dependent energy barriers
\begin{equation}
    E_\mathrm{IMT}(T)=\epsilon_\mathrm{IMT}(T_\mathrm{IMT}-T)~~~;~~~E_\mathrm{MIT}(T)=\epsilon_\mathrm{MIT}(T-T_\mathrm{MIT})~~,
\end{equation}
where $\epsilon_\mathrm{IMT}$ and $\epsilon_\mathrm{MIT}$ are the energy barrier change per kelvin, while $T_\mathrm{IMT}$ and $T_\mathrm{MIT}$ mark the temperature where the energy barriers for the IMT and MIT vanishes (for further details and the values see Section~2.2 and Table~1 of the ESI).
\section*{Acknowledgements}
This research was supported by the Ministry of Culture and Innovation and the National Research, Development and Innovation Office within the Quantum Information National Laboratory of Hungary (Grant No. 2022-2.1.1-NL-2022-00004), and the NKFI FK146339, K143169, K143282 grants. Z.B., T.N.T and L.P. acknowledge the support of the Bolyai J\'{a}nos Research Scholarship of the Hungarian Academy of Sciences.

\section*{Author contributions}
The VO$_2$ thin layers were grown by G.M and devices were fabricated by L.P. and T.N.T under the supervision of J.V. The noise measurements were performed by S.W.S under the supervison of Z.B. The high-frequency measurements were performed by S.W.S with contribution of B.S. L.P. developed the basics of two-dimensional resistor network model and S.W.S. complemented it by the simulation of noise. S.W.S. performed the numerical simulations and the fitting of the simulation results to the experiment. Cycle-to-cycle measurements for determining the threshold variations were carried out by Z.S.-S under the supervision of B.S. The project was supervised by A.H. and Z.B. The manuscript was prepared by S.W.S, Z.B, B.S. and A.H. All authors contributed to the discussion of the results.

\section*{Associated Content}
Electronic Supporting Information (ESI) is available, including details of the simulations, the determination of the model parameters, and the fitting of the simulation results to the experimental data. In addition, the ESI includes a simplified parallel-electrode model for intrinsic noise scaling in the LRS and a detailed analysis of the cycle-to-cycle variability of the threshold voltages.

\bibliography{references}

@article{Rocco2022,
   author = {Rodolfo Rocco and Javier del Valle and Henry Navarro and Pavel Salev and Ivan K. Schuller and Marcelo Rozenberg},
   doi = {10.1103/PhysRevApplied.17.024028},
   issn = {2331-7019},
   issue = {2},
   journal = {Physical Review Applied},
   month = {2},
   pages = {024028},
   title = {Exponential Escape Rate of Filamentary Incubation in Mott Spiking Neurons},
   volume = {17},
   year = {2022}
}

@article{Gao2022,
   author = {Xing Gao and Carlos M. M. Rosário and Hans Hilgenkamp},
   doi = {10.1063/5.0077160},
   issn = {2158-3226},
   issue = {1},
   journal = {AIP Advances},
   month = {1},
   title = {Multi-level operation in VO$_2$-based resistive switching devices},
   volume = {12},
   year = {2022}
}

@article{Maher2024,
   author = {Olivier Maher and Roy Bernini and Nele Harnack and Bernd Gotsmann and Marilyne Sousa and Valeria Bragaglia and Siegfried Karg},
   doi = {10.1038/s41598-024-61294-x},
   issn = {2045-2322},
   issue = {1},
   journal = {Scientific Reports},
   month = {5},
   pages = {11600},
   title = {Highly reproducible and CMOS-compatible VO$_2$-based oscillators for brain-inspired computing},
   volume = {14},
   year = {2024}
}

@article{Topalian2015,
   author = {Zareh Topalian and Shu-Yi Li and Gunnar A. Niklasson and Claes G. Granqvist and Laszlo B. Kish},
   doi = {10.1063/1.4905739},
   issn = {0021-8979},
   issue = {2},
   journal = {Journal of Applied Physics},
   month = {1},
   title = {Resistance noise at the metal–insulator transition in thermochromic VO$_2$ films},
   volume = {117},
   year = {2015}
}

@article{Gunes2024,
   author = {Ozan Gunes and Onyebuchi I. Onumonu and A. Baset Gholizadeh and Chunzi Zhang and Qiaoqin Yang and Shi-Jie Wen and Richard J. Curry and Robert E. Johanson and Safa O. Kasap},
   doi = {10.1063/5.0218097},
   issn = {0021-8979},
   issue = {1},
   journal = {Journal of Applied Physics},
   month = {7},
   title = {Excess noise and thermoelectric effect in magnetron-sputtered VO$_2$ thin films},
   volume = {136},
   year = {2024}
}

@article{Carapezzi2022,
   author = {Stefania Carapezzi and Corentin Delacour and Andrew Plews and Ahmed Nejim and Siegfried Karg and Aida Todri-Sanial},
   doi = {10.1038/s41598-022-23629-4},
   issn = {2045-2322},
   issue = {1},
   journal = {Scientific Reports},
   month = {11},
   pages = {19377},
   title = {Role of ambient temperature in modulation of behavior of vanadium dioxide volatile memristors and oscillators for neuromorphic applications},
   volume = {12},
   year = {2022}
}

@article{Jung2021,
   author = {Youngho Jung and Junho Jeong and Zhongnan Qu and Bin Cui and Ankita Khanda and Stuart S. P. Parkin and Joyce K. S. Poon},
   doi = {10.1002/aelm.202001142},
   issn = {2199-160X},
   issue = {8},
   journal = {Advanced Electronic Materials},
   month = {8},
   title = {Observation of Optically Addressable Nonvolatile Memory in VO$_2$ at Room Temperature},
   volume = {7},
   year = {2021}
}

@article{Haddad2022,
   author = {Emile Haddad and Roman V. Kruzelecky and Piotr Murzionak and Wes Jamroz and Kamel Tagziria and Mohamed Chaker and Boris Ledrogoff},
   doi = {10.3389/fmats.2022.1013848},
   issn = {2296-8016},
   journal = {Frontiers in Materials},
   month = {12},
   title = {Review of the VO$_2$ smart material applications with emphasis on its use for spacecraft thermal control},
   volume = {9},
   year = {2022}
}

@article{Li2022,
   author = {Ge Li and Donggang Xie and Hai Zhong and Ziye Zhang and Xingke Fu and Qingli Zhou and Qiang Li and Hao Ni and Jiaou Wang and Er-jia Guo and Meng He and Can Wang and Guozhen Yang and Kuijuan Jin and Chen Ge},
   doi = {10.1038/s41467-022-29456-5},
   issn = {2041-1723},
   issue = {1},
   journal = {Nature Communications},
   month = {4},
   pages = {1729},
   title = {Photo-induced non-volatile VO$_2$ phase transition for neuromorphic ultraviolet sensors},
   volume = {13},
   year = {2022}
}

@article{Ko2008,
   author = {Changhyun Ko and Shriram Ramanathan},
   doi = {10.1063/1.3050464},
   issn = {0003-6951},
   issue = {25},
   journal = {Applied Physics Letters},
   month = {12},
   title = {Observation of electric field-assisted phase transition in thin film vanadium oxide in a metal-oxide-semiconductor device geometry},
   volume = {93},
   year = {2008}
}

@article{Schmid2024,
   author = {Sebastian Werner Schmid and László Pósa and Tímea Nóra Török and Botond Sánta and Zsigmond Pollner and György Molnár and Yannik Horst and János Volk and Juerg Leuthold and András Halbritter and Miklós Csontos},
   doi = {10.1021/acsnano.4c03840},
   issn = {1936-0851},
   issue = {33},
   journal = {ACS Nano},
   month = {8},
   pages = {21966-21974},
   title = {Picosecond Femtojoule Resistive Switching in Nanoscale VO$_2$ Memristors},
   volume = {18},
   year = {2024}
}

@article{Psa2023,
   author = {László Pósa and Péter Hornung and Tímea Nóra Török and Sebastian Werner Schmid and Sadaf Arjmandabasi and György Molnár and Zsófia Baji and Goran Dražić and András Halbritter and János Volk},
   doi = {10.1021/acsanm.3c00150},
   issn = {2574-0970},
   issue = {11},
   journal = {ACS Applied Nano Materials},
   month = {6},
   pages = {9137-9147},
   title = {Interplay of Thermal and Electronic Effects in the Mott Transition of Nanosized VO$_2$ Phase Change Memory Devices},
   volume = {6},
   year = {2023}
}

@article{delValle2019,
   author = {Javier del Valle and Pavel Salev and Federico Tesler and Nicolás M. Vargas and Yoav Kalcheim and Paul Wang and Juan Trastoy and Min-Han Lee and George Kassabian and Juan Gabriel Ramírez and Marcelo J. Rozenberg and Ivan K. Schuller},
   doi = {10.1038/s41586-019-1159-6},
   issn = {0028-0836},
   issue = {7756},
   journal = {Nature},
   month = {5},
   pages = {388-392},
   title = {Subthreshold firing in Mott nanodevices},
   volume = {569},
   year = {2019}
}

@article{Son2012,
   author = {Myungwoo Son and Xinjun Liu and Sharif Md. Sadaf and Daeseok Lee and Sangsu Park and Wootae Lee and Seonghyun Kim and Jubong Park and Jungho Shin and Seungjae Jung and Moon-Ho Ham and Hyunsang Hwang},
   doi = {10.1109/LED.2012.2188989},
   issn = {0741-3106},
   issue = {5},
   journal = {IEEE Electron Device Letters},
   month = {5},
   pages = {718-720},
   title = {Self-Selective Characteristics of Nanoscale {VO}$_x$ Devices for High-Density ReRAM Applications},
   volume = {33},
   year = {2012}
}

@article{Psa2021,
   author = {László Pósa and György Molnár and Benjamin Kalas and Zsófia Baji and Zsolt Czigány and Péter Petrik and János Volk},
   doi = {10.3390/nano11010212},
   issn = {2079-4991},
   issue = {1},
   journal = {Nanomaterials},
   month = {1},
   pages = {212},
   title = {A Rational Fabrication Method for Low Switching-Temperature VO$_2$},
   volume = {11},
   year = {2021}
}

@article{Kalcheim2020,
   author = {Yoav Kalcheim and Alberto Camjayi and Javier del Valle and Pavel Salev and Marcelo Rozenberg and Ivan K. Schuller},
   doi = {10.1038/s41467-020-16752-1},
   issn = {2041-1723},
   issue = {1},
   journal = {Nature Communications},
   month = {6},
   pages = {2985},
   title = {Non-thermal resistive switching in Mott insulator nanowires},
   volume = {11},
   year = {2020}
}

@article{Qazilbash2007,
   author = {M. M. Qazilbash and M. Brehm and Byung-Gyu Chae and P.-C. Ho and G. O. Andreev and Bong-Jun Kim and Sun Jin Yun and A. V. Balatsky and M. B. Maple and F. Keilmann and Hyun-Tak Kim and D. N. Basov},
   doi = {10.1126/science.1150124},
   issn = {0036-8075},
   issue = {5857},
   journal = {Science},
   month = {12},
   pages = {1750-1753},
   title = {Mott Transition in VO$_2$ Revealed by Infrared Spectroscopy and Nano-Imaging},
   volume = {318},
   year = {2007}
}

@article{Sohn2015,
   author = {Ahrum Sohn and Teruo Kanki and Kotaro Sakai and Hidekazu Tanaka and Dong-Wook Kim},
   doi = {10.1038/srep10417},
   issn = {2045-2322},
   issue = {1},
   journal = {Scientific Reports},
   month = {5},
   pages = {10417},
   title = {Fractal Nature of Metallic and Insulating Domain Configurations in a VO$_2$ Thin Film Revealed by Kelvin Probe Force Microscopy},
   volume = {5},
   year = {2015}
}

@article{Balogh2021,
   author = {Zoltán Balogh and Gréta Mezei and László Pósa and Botond Sánta and András Magyarkuti and András Halbritter},
   doi = {10.1088/2399-1984/ac14c8},
   issn = {2399-1984},
   issue = {4},
   journal = {Nano Futures},
   month = {12},
   pages = {042002},
   title = {1/f noise spectroscopy and noise tailoring of nanoelectronic devices},
   volume = {5},
   year = {2021}
}

@article{Kim2025,
   author = {Han Gyeol Kim and Deok Hun Kim and Jehoon Lee and Junyeob Yeo and Joonghoe Dho},
   doi = {10.1016/j.mssp.2025.109276},
   issn = {13698001},
   journal = {Materials Science in Semiconductor Processing},
   month = {4},
   pages = {109276},
   title = {Resistive and photocurrent switching behaviors of a flexible VO$_2$/mica device fabricated via laser ablation patterning},
   volume = {189},
   year = {2025}
}

@article{Wu2006,
   author = {Junqiao Wu and Qian Gu and Beth S. Guiton and Nathalie P. de Leon and Lian Ouyang and Hongkun Park},
   doi = {10.1021/nl061831r},
   issn = {1530-6984},
   issue = {10},
   journal = {Nano Letters},
   month = {10},
   pages = {2313-2317},
   title = {Strain-Induced Self Organization of Metal-Insulator Domains in Single-Crystalline VO$_2$ Nanobeams},
   volume = {6},
   year = {2006}
}

@Book{Kogan1996,
  title     = {Electronic Noise and Fluctuations in Solids},
  publisher = {Cambridge University Press, New York},
  year      = {1996},
  author    = {Sh. Kogan},
}

@article{cahn1958free,
  title={Free energy of a nonuniform system. I. Interfacial free energy},
  author={Cahn, John W and Hilliard, John E},
  journal={The Journal of chemical physics},
  volume={28},
  number={2},
  pages={258--267},
  year={1958},
  publisher={American Institute of Physics}
}

@article{Tian2018,
	title = {Reconfigurable {Vanadium} {Dioxide} {Nanomembranes} and {Microtubes} with {Controllable} {Phase} {Transition} {Temperatures}},
	volume = {18},
	issn = {1530-6984},
	url = {https://doi.org/10.1021/acs.nanolett.8b00483},
	doi = {10.1021/acs.nanolett.8b00483},
	number = {5},
	urldate = {2026-04-21},
	journal = {Nano Letters},
	publisher = {American Chemical Society},
	author = {Tian, Ziao and Xu, Borui and Hsu, Bo and Stan, Liliana and Yang, Zheng and Mei, YongFeng},
	month = may,
	year = {2018},
	pages = {3017--3023},
}

@article{Cao2009,
	title = {Strain engineering and one-dimensional organization of metal–insulator domains in single-crystal vanadium dioxide beams},
	volume = {4},
	copyright = {2009 Springer Nature Limited},
	issn = {1748-3395},
	url = {https://www.nature.com/articles/nnano.2009.266},
	doi = {10.1038/nnano.2009.266},
	language = {en},
	number = {11},
	urldate = {2026-04-21},
	journal = {Nature Nanotechnology},
	publisher = {Nature Publishing Group},
	author = {Cao, J. and Ertekin, E. and Srinivasan, V. and Fan, W. and Huang, S. and Zheng, H. and Yim, J. W. L. and Khanal, D. R. and Ogletree, D. F. and Grossman, J. C. and Wu, J.},
	month = nov,
	year = {2009},
	pages = {732--737},
}

@article{Fang2022,
	title = {A bioinspired flexible artificial mechanoreceptor based on {VO$_2$} insulator-metal transition memristor},
    journal = {Journal of Alloys and Compounds},
	author = {Fang, Sheng Li and Han, Chuan Yu and Liu, Weihua and Han, Zheng Rong and Ma, Bo and Cui, Yi Lin and Fan, Shi Quan and Li, Xin and Wang, Xiao Li and Zhang, Guo He and Yin, Jun Qing and Huang, Xiao Dong and Geng, Li},
	volume = {911},
	issn = {0925-8388},
	url = {https://www.sciencedirect.com/science/article/pii/S0925838822014876},
	doi = {10.1016/j.jallcom.2022.165096},
	month = aug,
	year = {2022},
	pages = {165096},
}

@Article{Dutta2021,
author={Dutta, S.
and Khanna, A.
and Assoa, A. S.
and Paik, H.
and Schlom, D. G.
and Toroczkai, Z.
and Raychowdhury, A.
and Datta, S.},
title={An Ising Hamiltonian solver based on coupled stochastic phase-transition nano-oscillators},
journal={Nature Electronics},
year={2021},
month={Jul},
day={01},
volume={4},
number={7},
pages={502-512},
issn={2520-1131},
doi={10.1038/s41928-021-00616-7}
}

@Article{Mohseni2022,
author={Mohseni, Naeimeh
and McMahon, Peter L.
and Byrnes, Tim},
title={Ising machines as hardware solvers of combinatorial optimization problems},
journal={Nature Reviews Physics},
year={2022},
month={Jun},
day={01},
volume={4},
number={6},
pages={363-379},
issn={2522-5820},
doi={10.1038/s42254-022-00440-8}
}

@article{Yi2018,
	title = {Biological plausibility and stochasticity in scalable VO$_2$ active memristor neurons},
	volume = {9},
	doi = {10.1038/s41467-018-07052-w},
	journal = {Nature Communications},
	author = {Yi, Wei and Tsang, Kenneth and Lam, Stephen and Bai, Xiwei and Crowell, Jack and Flores, Elias},
	month = nov,
	year = {2018},
	pages = {4661},
}

@Article{Gopalakrishnan2009,
author={Gopalakrishnan, Gokul
and Ruzmetov, Dmitry
and Ramanathan, Shriram},
title={On the triggering mechanism for the metal--insulator transition in thin film VO$_2$ devices: electric field versus thermal effects},
journal={Journal of Materials Science},
year={2009},
month={Oct},
day={01},
volume={44},
number={19},
pages={5345-5353},
issn={1573-4803},
doi={10.1007/s10853-009-3442-7},
url={https://doi.org/10.1007/s10853-009-3442-7}
}

@article{Tiwari2026,
	title = {Near field optical visualization of the nanoscale phase percolation dynamics of a {VO$_2$} oscillator},
	volume = {17},
	copyright = {2026 The Author(s)},
	issn = {2041-1723},
	url = {https://www.nature.com/articles/s41467-026-68300-y},
	doi = {10.1038/s41467-026-68300-y},
	language = {en},
	number = {1},
	urldate = {2026-04-21},
	journal = {Nature Communications},
	publisher = {Nature Publishing Group},
	author = {Tiwari, Kajal and Wang, Zhong and Xie, Yishen and Gopi, Ajesh Kollakuzhiyil and Jeon, Jae-Chun and Xiao, Ke and Parkin, Stuart S. P.},
	month = jan,
	year = {2026},
	pages = {600},
}

@article{Stinson2018,
	title = {Imaging the nanoscale phase separation in vanadium dioxide thin films at terahertz frequencies},
	volume = {9},
	copyright = {2018 The Author(s)},
	issn = {2041-1723},
	url = {https://www.nature.com/articles/s41467-018-05998-5},
	doi = {10.1038/s41467-018-05998-5},
	language = {en},
	number = {1},
	urldate = {2026-04-21},
	journal = {Nature Communications},
	publisher = {Nature Publishing Group},
	author = {Stinson, H. T. and Sternbach, A. and Najera, O. and Jing, R. and Mcleod, A. S. and Slusar, T. V. and Mueller, A. and Anderegg, L. and Kim, H. T. and Rozenberg, M. and Basov, D. N.},
	month = sep,
	year = {2018},
	pages = {3604},
}

@article{chandrashekhar1973heat,
  title={Heat capacity of VO$_2$ single crystals},
  author={Chandrashekhar, GV and Barros, HLC and Honig, JM},
  journal={Materials Research Bulletin},
  volume={8},
  number={4},
  pages={369--374},
  year={1973},
  publisher={Elsevier}
}

@online{VO2density,
title = {VO$_2$},
url = {https://www.chembk.com/en/chem/VO2},
urldate = {2023-08-16}
}

@article{QI20082300,
title = {Synthesis, characterization, and thermodynamic parameters of vanadium dioxide},
journal = {Materials Research Bulletin},
volume = {43},
number = {8},
pages = {2300-2307},
year = {2008},
issn = {0025-5408},
doi = {https://doi.org/10.1016/j.materresbull.2007.08.016},
url = {https://www.sciencedirect.com/science/article/pii/S0025540807003790},
author = {Ji Qi and Guiling Ning and Yuan Lin}
}

@article{li2021thermal,
  title={Thermal conductivity of VO$_2$ nanowires at metal-insulator transition temperature},
  author={Li, Da and Wang, Qilang and Xu, Xiangfan},
  journal={Nanomaterials},
  volume={11},
  number={9},
  pages={2428},
  year={2021},
  publisher={MDPI}
}

@article{oh2010thermal,
  title={Thermal conductivity and dynamic heat capacity across the metal-insulator transition in thin film VO$_2$},
  author={Oh, Dong-Wook and Ko, Changhyun and Ramanathan, Shriram and Cahill, David G},
  journal={Applied Physics Letters},
  volume={96},
  number={15},
  year={2010},
  publisher={AIP Publishing}
}

@article{santos2013thermoelectric,
  title={Thermoelectric properties of V2O5 thin films deposited by thermal evaporation},
  author={Santos, R and Loureiro, J and Nogueira, A and Elangovan, E and Pinto, JV and Veiga, JP and Busani, T and Fortunato, E and Martins, R and Ferreira, I},
  journal={Applied Surface Science},
  volume={282},
  pages={590--594},
  year={2013},
  publisher={Elsevier}
}

@article{wang2020thermal,
  title={Thermal conductivity of V 2 O 5 nanowires and their contact thermal conductance},
  author={Wang, Qilang and Liang, Xing and Liu, Bohai and Song, Yihui and Gao, Guohua and Xu, Xiangfan},
  journal={Nanoscale},
  volume={12},
  number={2},
  pages={1138--1143},
  year={2020},
  publisher={Royal Society of Chemistry}
}

@article{andreev1980low,
  title={Low-frequency noise in vanadium dioxide undergoing a metal--semiconductor phase transition},
  author={Andreev, VN and Zakharchenya, BP and Kapshin, Yu S and Noskin, VA and Chudnovskil, FA},
  journal={The Journal of Experimental and Theoretical Physics (English translation)},
  volume={79},
  pages={1353},
  year={1980}
}

@article{almeida2000,
  title={Thermal dynamics of VO$_2$ films within the metal--insulator transition: Evidence for chaos near percolation threshold},
  author={De Almeida, LAL and Deep, GS and Lima, AMN and Neff, H},
  journal={Applied Physics Letters},
  volume={77},
  number={26},
  pages={4365--4367},
  year={2000},
  publisher={American Institute of Physics}
}

@article{baidakova1997structural,
  title={Structural and noise characterization of VO$_2$ films on SiO$_2$/Si substrates},
  author={Baidakova, MV and Bobyl’, AV and Malyarov, VG and Tret’yakov, VV and Khrebtov, IA and Shaganov, II},
  journal={Technical Physics Letters},
  volume={23},
  number={7},
  pages={520--522},
  year={1997},
  publisher={Springer}
}

@article{basantani2012enhanced,
  title={Enhanced electrical and noise properties of nanocomposite vanadium oxide thin films by reactive pulsed-dc magnetron sputtering},
  author={Basantani, HA and Kozlowski, S and Lee, Myung-Yoon and Li, J and Dickey, EC and Jackson, TN and Bharadwaja, SSN and Horn, M},
  journal={Applied Physics Letters},
  volume={100},
  number={26},
  year={2012},
  publisher={AIP Publishing}
}

@article{jelks1975response,
  title={Response of thermal filaments in VO$_2$ to laser- produced thermal perturbations},
  author={Jelks, EC and Walser, RM and Ben{\'e}, RW and Neal II, WH},
  journal={Applied Physics Letters},
  volume={26},
  number={7},
  pages={355--357},
  year={1975},
  publisher={AIP Publishing}
}

@article{velichko2003deterministic,
  title={Deterministic noise in vanadium dioxide based structures},
  author={Velichko, AA and Stefanovich, GB and Pergament, AL and Boriskov, PP},
  journal={Technical Physics Letters},
  volume={29},
  number={5},
  pages={435--437},
  year={2003},
  publisher={Springer}
}

@article{morin1959oxides,
  title={Oxides which show a metal-to-insulator transition at the Neel temperature},
  author={Morin, FJ},
  journal={Physical review letters},
  volume={3},
  number={1},
  pages={34},
  year={1959},
  publisher={APS}
}

@article{goodenough1971two,
  title={The two components of the crystallographic transition in VO$_2$},
  author={Goodenough, John B},
  journal={Journal of Solid State Chemistry},
  volume={3},
  number={4},
  pages={490--500},
  year={1971},
  publisher={Elsevier}
}

@article{zylbersztejn1975metal,
  title={Metal-insulator transition in vanadium dioxide},
  author={Zylbersztejn, AMNF and Mott, Nevill Francis},
  journal={Physical Review B},
  volume={11},
  number={11},
  pages={4383},
  year={1975},
  publisher={APS}
}

@article{kiss1993new,
  title={New noise exponents in random conductor-superconductor and conductor-insulator mixtures},
  author={Kiss, LB and Svedlindh, P},
  journal={Physical review letters},
  volume={71},
  number={17},
  pages={2817},
  year={1993},
  publisher={APS}
}

@article{hui1986noise,
  title={Noise exponent in superconducting-normal metal mixtures},
  author={Hui, PM and Stroud, D},
  journal={Physical Review B},
  volume={34},
  number={11},
  pages={8101},
  year={1986},
  publisher={APS}
}

@article{pollner2026vo2,
  title={VO$_2$ Oscillator Circuits Optimized for Ultrafast, 100 MHz-Range Operation},
  author={Pollner, Zsigmond and T{\"o}r{\"o}k, T{\'\i}mea N{\'o}ra and P{\'o}sa, L{\'a}szl{\'o} and Csontos, Mikl{\'o}s and Schmid, Sebastian Werner and Balogh, Zolt{\'a}n and B{\"u}kkfejes, Andr{\'a}s and Kim, Heungsoo and Piqu{\'e}, Alberto and Leuthold, Juerg and others},
  journal={Advanced Electronic Materials},
  volume={12},
  number={1},
  pages={e00433},
  year={2026},
  publisher={Wiley Online Library}
}

@article{molnar2026neural,
  title={Neural Information Processing and Time-Series Prediction with Only Two Dynamical Memristors},
  author={Moln{\'a}r, D{\'a}niel and T{\"o}r{\"o}k, T{\'\i}mea N{\'o}ra and Volk, J{\'a}nos and K{\"o}vecs, Roland and P{\'o}sa, L{\'a}szl{\'o} and Bal{\'a}zs, P{\'e}ter and Moln{\'a}r, Gy{\"o}rgy and Olalla, Nadia Jimenez and Balogh, Zolt{\'a}n and Volk, J{\'a}nos and others},
  journal={Advanced Electronic Materials},
  volume={12},
  number={7},
  pages={e00353},
  year={2026},
  publisher={Wiley Online Library}
}

@article{santa2021noise,
  title={Noise tailoring in memristive filaments},
  author={Santa, Botond and Balogh, Zoltan and Posa, Laszlo and Krisztian, David and Torok, Timea Nora and Molnar, Daniel and Sinko, Csaba and Hauert, Roland and Csontos, Miklos and Halbritter, Andras},
  journal={ACS applied materials \& interfaces},
  volume={13},
  number={6},
  pages={7453--7460},
  year={2021},
  publisher={ACS Publications}
}

@article{santa2019universal,
  title={Universal 1/f type current noise of Ag filaments in redox-based memristive nanojunctions},
  author={S{\'a}nta, Botond and Balogh, Zolt{\'a}n and Gubicza, Agnes and P{\'o}sa, L{\'a}szl{\'o} and Kriszti{\'a}n, D{\'a}vid and Mih{\'a}ly, Gy{\"o}rgy and Csontos, Mikl{\'o}s and Halbritter, Andr{\'a}s},
  journal={Nanoscale},
  volume={11},
  number={11},
  pages={4719--4725},
  year={2019},
  publisher={Royal Society of Chemistry}
}

@article{balogh2023configuration,
  title={Configuration-specific insight into single-molecule conductance and noise data revealed by the principal component projection method},
  author={Balogh, Zolt{\'a}n and Mezei, Gr{\'e}ta and Tenk, N and Magyarkuti, Andr{\'a}s and Halbritter, A},
  journal={The Journal of Physical Chemistry Letters},
  volume={14},
  number={22},
  pages={5109--5118},
  year={2023},
  publisher={ACS Publications}
}

@article{posa2021noise,
  title={Noise diagnostics of graphene interconnects for atomic-scale electronics},
  author={P{\'o}sa, L{\'a}szl{\'o} and Balogh, Zolt{\'a}n and Kriszti{\'a}n, D{\'a}vid and Bal{\'a}zs, P{\'e}ter and S{\'a}nta, Botond and Furrer, Roman and Csontos, Miklos and Halbritter, Andr{\'a}s},
  journal={npj 2D Materials and Applications},
  volume={5},
  number={1},
  pages={57},
  year={2021},
  publisher={Nature Publishing Group UK London}
}

@article{nyary2025benchmarking,
  title={Benchmarking stochasticity behind reproducibility: denoising strategies in Ta$_2$O$_5$ memristors},
  author={Ny{\'a}ry, Anna and Balogh, Zolt{\'a}n and S{\'a}nta, Botond and Lazar, Gyorgy and Jimenez Olalla, Nadia and Leuthold, Juerg and Csontos, Mikl{\'o}s and Halbritter, Andr{\'a}s},
  journal={ACS Applied Materials \& Interfaces},
  volume={17},
  number={17},
  pages={25654--25662},
  year={2025},
  publisher={ACS Publications}
}

@article{qazilbash2008electrodynamics,
  title={Electrodynamics of the vanadium oxides VO$_2$ and V$_2$O$_3$},
  author={Qazilbash, Mumtaz M and Schafgans, AA and Burch, KS and Yun, SJ and Chae, BG and Kim, BJ and Kim, Hyun-Tak and Basov, DN},
  journal={Physical Review B—Condensed Matter and Materials Physics},
  volume={77},
  number={11},
  pages={115121},
  year={2008},
  publisher={APS}
}

@article{belozerov2011evidence,
  title={Evidence for strong Coulomb correlations in the metallic phase of vanadium dioxide},
  author={Belozerov, Aleksandr Sergeevich and Poteryaev, Alexandr Ivanovich and Anisimov, Vladimir Il'ich},
  journal={JETP letters},
  volume={93},
  number={2},
  pages={70--74},
  year={2011},
  publisher={Springer}
}

@article{Ahn2021,
author = {Ahn, Heebeom and Kang, Keehoon and Song, Younggul and Lee, Woocheol and Kim, Jae-Keun and Kim, Junwoo and Lee, Jonghoon and Baek, Kyeong-Yoon and Shin, Jiwon and Lim, Hyungbin and Kim, Yongjin and Lee, Jae Sung and Lee, Takhee},
title = {Resistive Switching by Percolative Conducting Filaments in Organometal Perovskite Unipolar Memory Devices Analyzed Using Current Noise Spectra},
journal = {Advanced Functional Materials},
volume = {32},
number = {4},
pages = {2107727},
doi = {https://doi.org/10.1002/adfm.202107727},
url = {https://advanced.onlinelibrary.wiley.com/doi/abs/10.1002/adfm.202107727},
eprint = {https://advanced.onlinelibrary.wiley.com/doi/pdf/10.1002/adfm.202107727},
year = {2022}
}

@article{Fehervari2024,
    author = {Fehérvári, János Gergő and Balogh, Zoltán and Török, Tímea Nóra and Halbritter, András},
    title = {Noise tailoring, noise annealing, and external perturbation injection strategies in memristive Hopfield neural networks},
    journal = {APL Machine Learning},
    volume = {2},
    number = {1},
    pages = {016107},
    year = {2024},
    month = {01},
    issn = {2770-9019},
    doi = {10.1063/5.0173662},
    url = {https://doi.org/10.1063/5.0173662},
    eprint = {https://pubs.aip.org/aip/aml/article-pdf/doi/10.1063/5.0173662/18704021/016107\_1\_5.0173662.pdf},
}

@INPROCEEDINGS{Lombardo2024,
  author={Lombardo, Davide G. F. and Ram, Mamidala Saketh and Stecconi, Tommaso and Choi, Wooseok and Porta, Antonio La and Falcone, Donato F. and Offrein, Bert and Bragaglia, Valeria},
  booktitle={2024 Device Research Conference (DRC)}, 
  title={Read Noise Analysis in Analog Conductive-Metal-Oxide/HfOx ReRAM Devices}, 
  year={2024},
  volume={},
  number={},
  pages={1-2},
  doi={10.1109/DRC61706.2024.10643760}}

@INPROCEEDINGS{Puglisi2016,
  author={Puglisi, Francesco Maria and Pavan, Paolo and Larcher, Luca},
  booktitle={2016 IEEE International Reliability Physics Symposium (IRPS)}, 
  title={Random telegraph noise in HfOx Resistive Random Access Memory: From physics to compact modeling}, 
  year={2016},
  volume={},
  number={},
  pages={MY-8-1-MY-8-5},
  doi={10.1109/IRPS.2016.7574624}}

@ARTICLE{Lee2011,
  author={Lee, Daeseok and Lee, Joonmyoung and Jo, Minseok and Park, Jubong and Siddik, Manzar and Hwang, Hyunsang},
  journal={IEEE Electron Device Letters}, 
  title={Noise-Analysis-Based Model of Filamentary Switching ReRAM With $\hbox{ZrO}_{x}/\hbox{HfO}_{x}$ Stacks}, 
  year={2011},
  volume={32},
  number={7},
  pages={964-966},
  doi={10.1109/LED.2011.2148689}}

@article{Rao2023,
   author = {Mingyi Rao and Hao Tang and Jiangbin Wu and Wenhao Song and Max Zhang and Wenbo Yin and Ye Zhuo and Fatemeh Kiani and Benjamin Chen and Xiangqi Jiang and Hefei Liu and Hung-Yu Chen and Rivu Midya and Fan Ye and Hao Jiang and Zhongrui Wang and Mingche Wu and Miao Hu and Han Wang and Qiangfei Xia and Ning Ge and Ju Li and J. Joshua Yang},
   doi = {10.1038/s41586-023-05759-5},
   issn = {0028-0836},
   issue = {7954},
   journal = {Nature},
   month = {3},
   pages = {823-829},
   title = {Thousands of Conductance Levels in Memristors Integrated on CMOS},
   volume = {615},
   url = {https://www.nature.com/articles/s41586-023-05759-5},
   year = {2023},
}
\bibliographystyle{achemso.bst}
\end{document}


\begin{center}
\noindent\LARGE{\textbf{Supporting information\\Voltage-driven transition from steady-state fluctuations to phase-transition noise in nanoscale VO$_2$ devices}}\\
\vspace{0.6cm}
\noindent\large{Sebastian Werner Schmid\textit{$^{a,b}$}, Zolt\'an Balogh$^{\ast}$\textit{$^{a,c}$}, Botond S\'anta \textit{$^{a}$}, T\'imea N\'ora Török\textit{$^{a,d}$}, Zsombor Sin\'oros-Szab\'o\textit{$^{a}$}, Gy\"orgy Moln\'ar\textit{$^{d}$}, J\'anos Volk \textit{$^{d}$}, L\'aszl\'o P\'osa \textit{$^{a,d}$}, and Andr\'as Halbritter\textit{$^{a,c}$}}\vspace{0.6cm}
\end{center}
\textit{$^{a}$~Department of Physics, Institute of Physics, Budapest University of Technology and Economics, Műegyetem rkp. 3., H-1111 Budapest, Hungary}\\
\textit{$^{b}$~Experimental Physics V, Center for Electronic Correlations and Magnetism, University of Augsburg, Augsburg 86159, Germany}\\
\textit{$^{c}$~HUN-REN-BME Condensed Matter Research Group, Műegyetem rkp. 3., H-1111 Budapest, Hungary}\\
\textit{$^{d}$~Institute of Technical Physics and Materials Science, HUN-REN Centre for Energy Research, Konkoly-Thege M. \'ut 29-33, H-1121 Budapest, Hungary}\\

$^{\ast}$\textit{Corresponding author: balogh.zoltan@ttk.bme.hu}\vspace{0.6cm}

\section{Details of the simulation}

\subsection{Basics of two-dimensional resistor network model}\label{sim_basics}
\begin{figure}[h!]
    \centering
    \includegraphics[width=0.6\columnwidth]{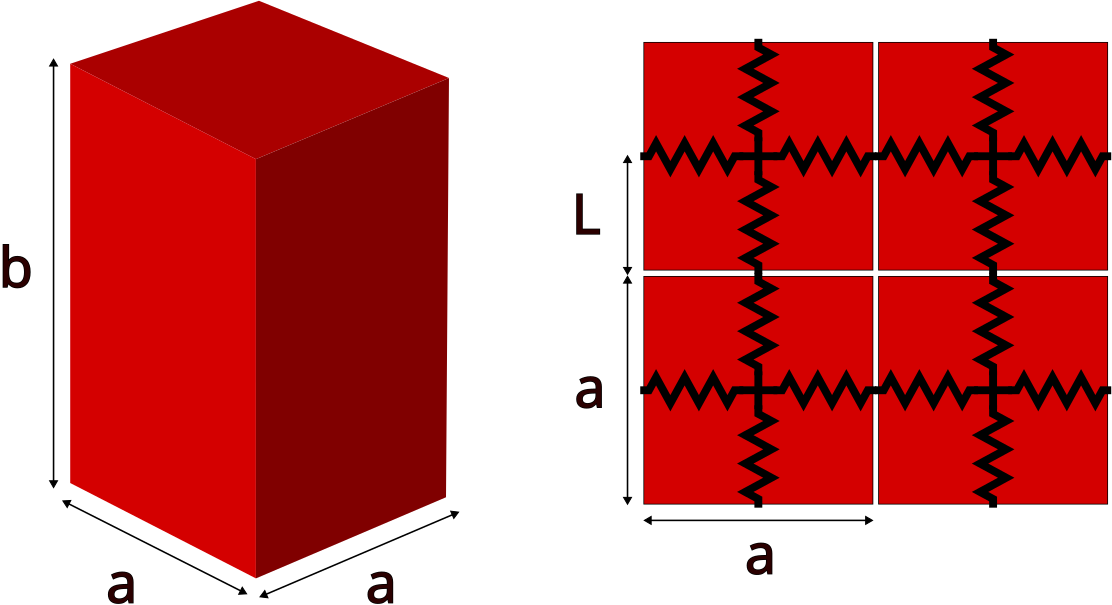}
    \caption{\textbf{2-D resistor network structure} Individual cells (left) of the two-dimensional resistor network simulation from the network (right) via coupling of their thermal and electrical resistances. Using the length $a$, width $a$ and height $b$, the intrinsic properties of VO$_2$ can be converted to the resistance, heat capacity and thermal conductance of a single cell. While the height $b$ is only a scaling parameter for the conversion of physical properties for one cell, the length and width are both equal to a and scales the size of the simulated network to the geometry of the real-world thin film.}
    \label{fig: network}
\end{figure}
The two-dimensional resistor network model consists of individual cells, that are aligned with each other, forming a two-dimensional layer as shown in Figure \ref{fig: network}. The length/width $a=5$~nm allows us to directly scale the size of the network to the size of the real-world device and replicate its structure. While the network forms a 2-D system, we attributed each cell the height of $b=40$~nm (thickness of our vanadium dioxide sample) to convert intrinsic physical properties to simulation parameters. The heat capacity per cell is then calculated by
\begin{equation}
    C=a^2bc_s\rho~,
\end{equation}
where $c_s\left[\frac{\mathrm{J}}{\mathrm{kg\cdot K}}\right]$ is the specific heat capacity and $\rho\left[\frac{\mathrm{kg}}{\mathrm{m}^3}\right]$ is the density. From the intrinsic thermal conductivity $\kappa\left[\frac{\mathrm{W}}{\mathrm{m\cdot K}}\right]$, we get the thermal conductance
\begin{equation}\label{eq: thermal conductance single cell}
    G=\frac{\kappa A}{L}=\frac{\kappa ab}{\frac{1}{2}a}=2\kappa b
\end{equation}
of one cell, where $A=ab$ is the interface area and $L=\frac{a}{2}$ the distance from its core to the interface. The thermal conductance between two cells $i$ and $j$, which are either in HRS or LRS, is then calculated as
\begin{equation}
    G_{ij}=\frac{G_iG_j}{G_i+G_j}=\frac{4\kappa_i\kappa_jb^2}{2(\kappa_i+\kappa_j)b}=\frac{2\kappa_i\kappa_j b}{\kappa_i+\kappa_j}~~,
\end{equation}
which is then used to calculate the heat transport per timestep of the simulation:
\begin{equation}
    \frac{\Delta Q_{ij}}{\Delta t}=G_{ij}\Delta T_{ij}
\end{equation}
For $\kappa_i=\kappa_j$ between two similar cells, this reduces to $G_{ij}=\kappa b$, the same as Equation \ref{eq: thermal conductance single cell} when $L=a$ for the distance of one cell core to the neighboring cell core, while the thermal transport between different cell types (insulating and metallic VO$_2$) is limited by the lower thermal conductivity of the insulating phase. Using the relation $\Delta T=\frac{\Delta Q}{C}$ and including Joule heat as well as the thermal transport to the substrate and all nearest neighbors, the total temperature change for a cell during one timestep is:
\begin{equation}\label{eq: temp change}
    \frac{\Delta T_i}{\Delta t}=\frac{P_i}{C_i}-\frac{G_s}{C_i}(T_i-T_0)-\frac{1}{C_i}\sum_{j}^{NN}G_{ij}(T_i-T_j)~~.
\end{equation}
Here, $G_s$ is the thermal conductance to the substrate and $T_0$ is the base temperature of the substrate. The Joule heating $P_i=I_i^2R_i$ is calculated from the local current $I_i$ through the resistance
\begin{equation}\label{eq: cell resistance}
    R_i(T,E)=R_i(T_0,0) \cdot \exp\left( \frac{E_{\mathrm{g},i}}{2 \, \mathrm{k}_\mathrm{B} T_i}- \frac{E_{\mathrm{g},i}}{2 \, \mathrm{k}_\mathrm{B} T_0} - \frac{E_i}{E_{\mathrm{c},i}} \right)~,
\end{equation}
where $R_i(T_0,0)$ is the cell resistance at the base temperature of the substrate $T_0$ and zero electric field $E$, $E_g$ the activation gap energy and $E_c$ the critical electric field. It is important to note that Equation \ref{eq: temp change} leads to the stability criterion $\Delta t<\frac{C_\mathrm{i}}{G}=\tau_\mathrm{Thermal}$ (simulation timestep must be smaller than the thermal time-constant) in order to prevent the local cell temperatures from oscillating and/or diverging in the simulation. Furthermore, we choose an even smaller timestep \mbox{$\Delta t=0.5~\mathrm{ps}\ll\tau_\mathrm{Thermal}\approx10~\mathrm{ps}$} to prevent large temperature changes between simulation steps.

\subsection{Implementation of noise}
Within the simulations, two sources of noise were introduced. First, the intrinsic cell resistance
\begin{equation}
    R(T,E,x) = R(T,E)\cdot(1+x\mathcal{N})
\end{equation}
fluctuates around its mean value $\langle R(T,E,x)\rangle=R(T,E)$ during each simulation step with a fluctuation strength $x$ and a random number distribution $\mathcal{N}$ with $\langle \mathcal{N}\rangle=0$ and $\langle \mathcal{N}^2\rangle=1$. This mirrors the resistance fluctuation that dominates at low voltages. As
\begin{equation}
    \Delta R^2=\langle (R(T,E,x)-\langle R(T,E,x)\rangle)^2\rangle=\langle(R(T,E)x\mathcal{N})^2\rangle=R^2(T,E)x^2~~,
\end{equation}
the relative noise $\frac{\Delta R(T,E)}{R(T,E)}=x$ for individual cells is constant. The fluctuation strength $x$ can be chosen arbitrarily to scale the noise in the simulations to the measured noise signal for the HRS and LRS respectively and therefore functions as a fitting parameter. \\
The second noise source in the simulations is the transition of the vanadium dioxide cells between the insulating and metallic state via their transition probabilities
\begin{equation}
    P_\mathrm{IMT}=\exp\left(\frac{-E_\mathrm{IMT}(T)}{\mathrm{k}_\mathrm{B}T}\right)~~~;~~~P_\mathrm{MIT}=\exp\left(\frac{-E_\mathrm{MIT}(T)}{\mathrm{k}_\mathrm{B}T}\right)~~~,
\end{equation}
$P_\mathrm{IMT}$ and $P_\mathrm{MIT}$ are the probabilities for an insulating cell to switch to a metallic cell and a metallic cell to switch to an insulating cell, respectively. The energy barriers
\begin{equation}\label{eq: barriers}
    E_\mathrm{IMT}(T)=\epsilon_\mathrm{IMT}(T_\mathrm{IMT}-T)~~~;~~~E_\mathrm{MIT}(T)=\epsilon_\mathrm{MIT}(T-T_\mathrm{MIT})~~,
\end{equation}
change by $\epsilon_\mathrm{IMT}$ ($\epsilon_\mathrm{MIT}$) per kelvin and becomes zero at $T_\mathrm{IMT}$ ($T_\mathrm{MIT}$), leading to transition probabilities $P_\mathrm{IMT}(T)=1$ for $T\ge T_\mathrm{IMT}$ and $P_\mathrm{MIT}(T)=1$ for $T\le T_\mathrm{MIT}$. In this model, neither $T_\mathrm{IMT}$ nor $T_\mathrm{MIT}$ is the actual transition temperature $T_\mathrm{c}=338$~K of VO$_2$, instead $E_\mathrm{IMT}(T=338~\mathrm{K})=E_\mathrm{MIT}(T=338~\mathrm{K})$ and therefore $P_\mathrm{IMT}(T=338~\mathrm{K})=P_\mathrm{MIT}(T=338~\mathrm{K})$ define the phase transition, where both phases become equally likely. In contrast to the intrinsic noise of the cells, where a simple scaling parameter $x$ describes the noise amplitude, here the resistance difference between the insulating and metallic cell as well as their respective switching probabilities, define the noise amplitude of a cell. While the resistance of the two different cell types are determined from the resistance of the VO$_2$ device, the energy-barriers and their resulting switching probabilities are determined from the experimental noise signals.

\section{Obtaining simulation parameters from experimental data}

Using model parameters that are consistent with realistic physical values and ensure matching between the simulation and the experiment is crucial for the reliability of the simulation. Thus, during the simulation, the parameters were determined in two ways: for some parameters, we used values from the literature, while for others the values were obtained by fitting the experimental data. The model parameters are summarized in Table~\ref{parameters}, along with their values used in our simulation, the method used to obtain them, and the literature values, where available. For the fitted parameters, we used the literature values as initial values. These model parameters can be divided into two groups: DC-like parameters (thermal parameters, electrical parameters) and parameters describing the dynamics. Accordingly, we used average values in one case and noise data in the other to determine their values, as described below.

\begin{table}[h!]
\centering
\begin{tabular}{|l|c|c|c|}
     \hline
     \textbf{Parameter}& \textbf{Value} & \textbf{Source} & \textbf{Literature value} \\ \hline
     $c_\mathrm{S}$& 690 J/kgK & literature& 600-800 J/kgK\cite{chandrashekhar1973heat,QI20082300} \\ \hline
     $\rho$& 4340 kg/m$^3$ & literature & 4339 kg/m$^3$ \cite{VO2density} \\ \hline
     $\kappa_\mathrm{HRS}$& 2.18 W/mK & literature+fit& 2.2-5.5 W/mK \cite{li2021thermal,oh2010thermal} \\ \hline
     $\kappa_\mathrm{LRS}$& 3.67 W/mK & literature+fit& 2.5-6 W/mK\cite{li2021thermal,oh2010thermal} \\ \hline
     $\kappa_\mathrm{substrate}$& 2.13 W/mK & literature+fit& 0.5-4 W/mK\cite{wang2020thermal,santos2013thermoelectric} \\ \hline
     $E_\mathrm{g,HRS}$& 460 meV & fit & \\ \hline
     $E_\mathrm{c,HRS}$& 1.5 MV/cm & fit& \\ \hline
     $E_\mathrm{g,LRS}$& 80 meV & fit&  \\ \hline
     $T_0$& 298~K &  & \\ \hline
     $R_\mathrm{HRS}(T_0)$& 118 kOhm & fit&  \\ \hline
     $R_\mathrm{LRS}(T_0)$& 349 Ohm & fit&  \\ \hline
     $\epsilon_\mathrm{IMT}$& 4.2 meV/K & fit&  \\ \hline
     $\epsilon_\mathrm{MIT}$& 3.2 meV/K & fit&  \\ \hline
     $T_\mathrm{IMT}$& 357 K & fit & \\ \hline
     $T_\mathrm{MIT}$& 313 K & fit & \\ \hline
\end{tabular}
\caption{The values of the parameters used in the simulation, together with their sources and literature values where available.}
\label{parameters}
\end{table}

\subsection{Determination of thermal and electrical parameters by fitting average experimental data}

\begin{figure}[h!]
    \centering
    \includegraphics[width=\columnwidth]{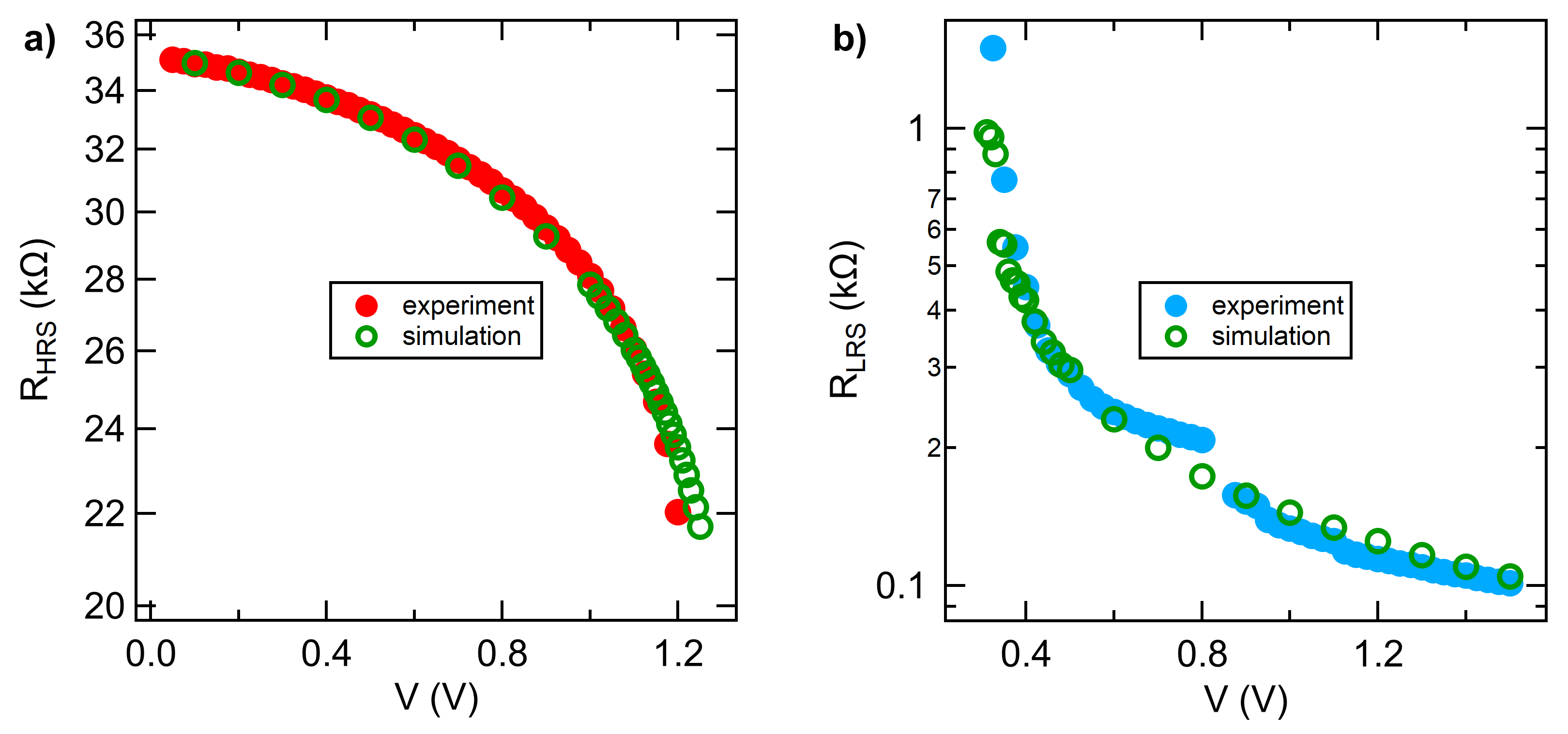}
    \caption{\textbf{experimental and simulated resistance versus voltage} Comparison of the experimental and simulated voltage dependent resistance of a) the high resistance state and b) the low resistance state.}
    \label{fig: resistance fits}
\end{figure}
After the geometry of the two-dimensional resistor network was set to that of the experimental device, the base resistance $R(T_0,0)$ (Equation \ref{eq: cell resistance}) can be determined for the high resistance state by matching the simulated resistance to the experimental resistance at $V\ll1$~V. The energy gaps $E_{g,\mathrm{HRS}}$ and $E_{g,\mathrm{LRS}}$ were extracted from the $R(T)$ data and the critical electric field $E_c$ by matching the slope in the simulations to the low voltage region of $R(V)$ where the temperature change from Joule heat ($\propto V^2$) is negligible. Then we adjusted the thermal parameters (starting values from literature) of the HRS so that the transition temperature $T_\mathrm{c}=338$~K is reached around the threshold voltages $V_\mathrm{set}=1.2$~V and $V_\mathrm{reset}=325$~mV of the experiments. The simulated $R(V)$ curve of the high resistance state is shown in comparison to the measured data in Figure \ref{fig: resistance fits}~a. \\
For the low resistance state, we concluded from the nanosecond timescale experiments (see Figure 4 of the main text) that there is no intrinsic electric field dependence of the resistance and therefore, we removed the electric field term in Equation \ref{eq: cell resistance} for the LRS. This leaves only a temperature dependence, originating from electron correlations in the metallic phase \cite{qazilbash2008electrodynamics,belozerov2011evidence}, which we approximate as a pseudo activation gap $E_\mathrm{g,LRS}$ for the LRS. While the experimental $R(T)$ data of the LRS close to $T_c$ can be described with $E_\mathrm{g,LRS}\approx200$~meV, this value only gives an upper limit for its simulation parameter, as at higher temperatures the metallic character of the LRS should become more pronounced. Getting the simulation parameters $R_\mathrm{LRS}(T_0,0)$, $\kappa_\mathrm{LRS}$ and $E_\mathrm{g,LRS}$ of the LRS was not as straightforward as for the HRS, as here the main contribution of the resistance change is from the volume changes of the metallic phase instead of the intrinsic cell resistance. The metallic volume depends on the local temperature, which, at a given voltage, is a result of the local cell resistance and the thermal conduction combined. This interplay between the parameters necessitated iterative adjustments to match the simulation curves to the experimental data, shown in Figure \ref{fig: resistance fits}~b.

\subsection{Matching simulated and experimental noise by adjusting dynamical parameters}
\begin{figure}[h!]
    \centering
    \includegraphics[width=0.8\columnwidth]{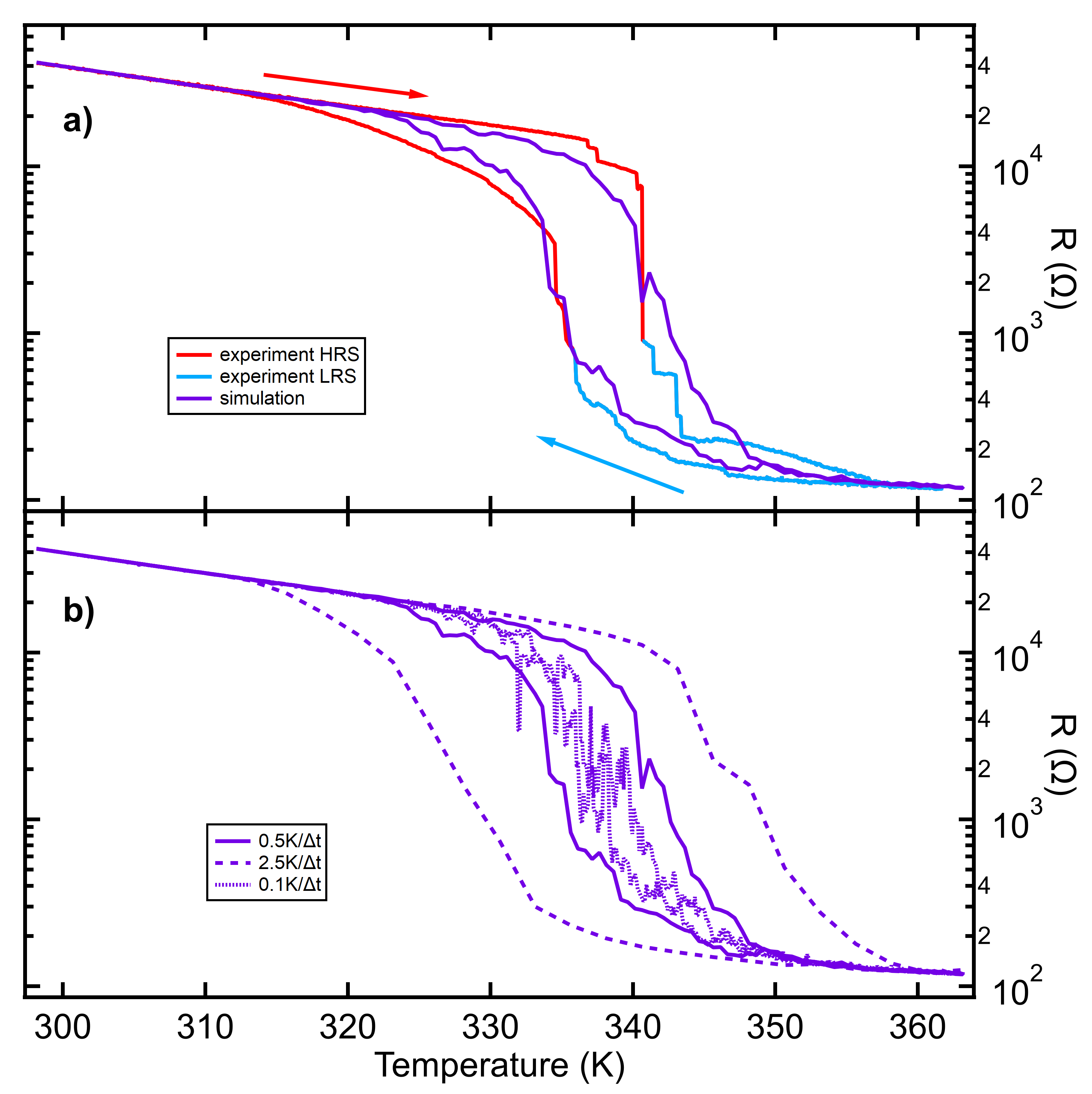}
    \caption{\textbf{experimental and simulated resistance versus temperature} a) Experimental and simulated temperature dependent resistance.b) Sweep rate dependence of the simulated $R(T)$ hysteresis curves.}
    \label{fig: temperature dependence}
\end{figure}
The transition between the insulating and metallic phase is determined by the energy barriers from Equation \ref{eq: barriers} which includes the four simulation parameters $\epsilon_\mathrm{IMT}$, $\epsilon_\mathrm{MIT}$, $T_\mathrm{IMT}$ and $T_\mathrm{MIT}$. First, we can estimate $T_\mathrm{IMT}\ge 357$~K and $T_\mathrm{MIT}\le 313$~K from the width of the experimental hysteresis region of Figure \ref{fig: temperature dependence}~a. We choose these parameters as the transition probability of $P_\mathrm{IMT}(T_\mathrm{IMT})=1$ ($P_\mathrm{MIT}(T_\mathrm{MIT})=1$) indicates the disappearance of the insulating (metallic) phase and therefore the end of the phase coexistence of the hysteresis region. Starting with the initial parameters $T_\mathrm{IMT}=357$~K and $T_\mathrm{MIT}=313$~K,
we then choose $\epsilon_\mathrm{IMT}$ as a fitting parameter, while $\epsilon_\mathrm{MIT}=\epsilon_\mathrm{IMT}\frac{T_\mathrm{IMT}-T_c}{T_c-T_\mathrm{MIT}}$ is fixed by the other parameters and the transition temperature $T_c=338$~K. A simulated $R(T)$ dataset for $\epsilon_\mathrm{IMT}=4.2$~meV/K and $\epsilon_\mathrm{MIT}=3.2$~meV/K is shown in Figure \ref{fig: temperature dependence}~a. While the simulated curve produces a hysteresis comparable to the experimental data, it is a highly time/sweep-rate dependent effect in the simulation as shown in Figure \ref{fig: temperature dependence}~b. There, an increased sweep rate from $0.5$~K/$\Delta$t (solid line) to $2.5$~K/$\Delta$t (dashed line) highly widens the hysteresis width, while a decreased sweep rate of $0.1$~K/$\Delta$t (dotted line) shrinks the hysteresis width, while leading to increased phase fluctuation close to $T_c$. As the simulated hysteresis is a function of the sweep rate as well as the timescale of the transition process, defined by the energy barriers, comparisons of $R(T)$-hysteresis traces with varying simulation parameters would necessitate excessive testing with various sweep rates for each set of simulation parameters and is therefore a suboptimal approach to compare simulation results.

\begin{figure}[h!]
    \centering
    \includegraphics[width=\columnwidth]{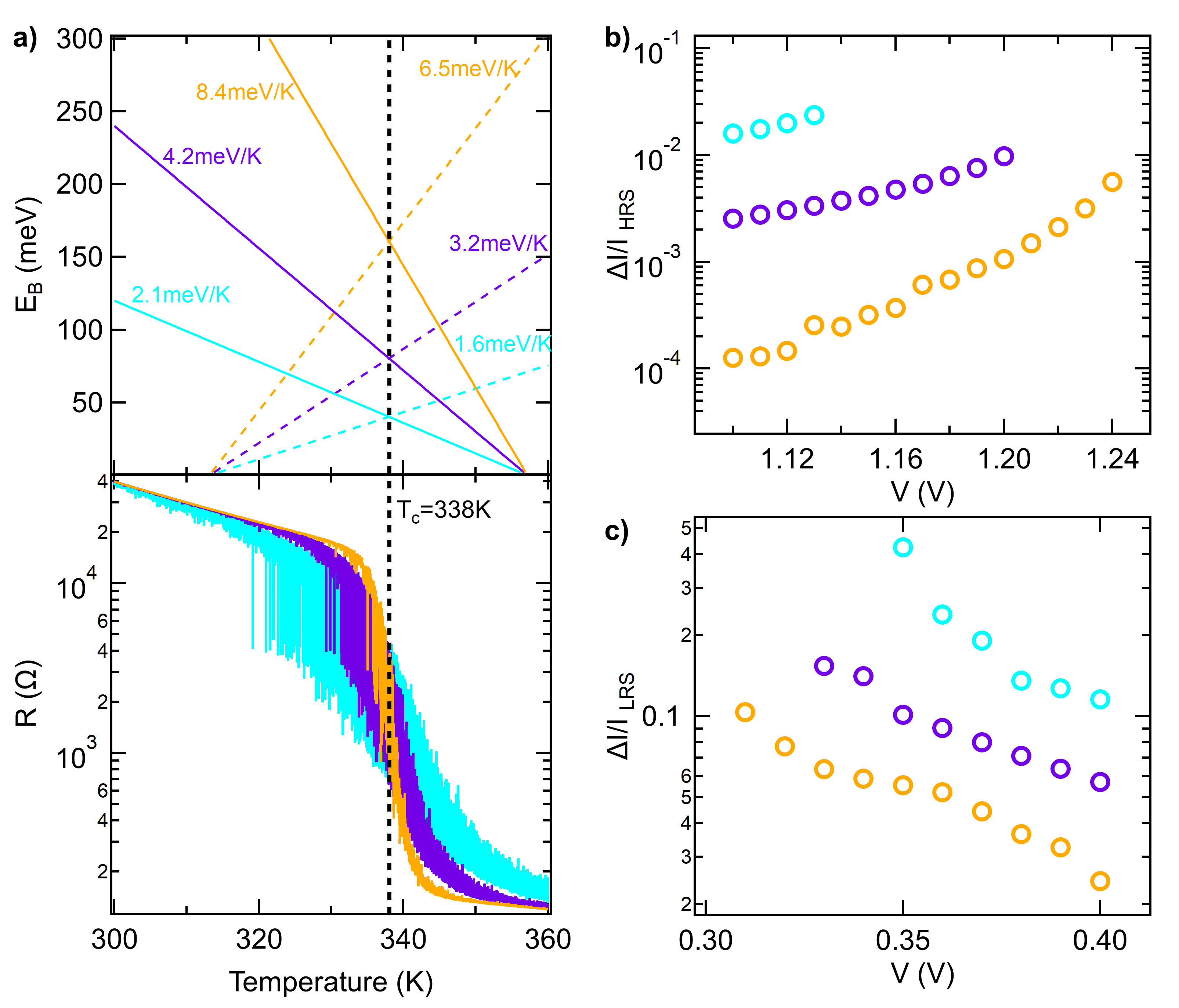}
    \caption{\textbf{Energy barrier change per Kelvin comparisons} a) Comparison of different energy barriers (upper panel) for the insulator metal transition (solid lines) with $\epsilon_\mathrm{IMT}=2.1$~meV/K (cyan), $\epsilon_\mathrm{IMT}=4.2$~meV/K (purple) and $\epsilon_\mathrm{IMT}=8.4$~meV/K (orange) for a fixed $T_\mathrm{IMT}=357$~K where the barrier becomes zero. The reset barriers for the metal to insulator transition (dashed lines) with $\epsilon_\mathrm{MIT}=1.6$~meV/K (cyan), $\epsilon_\mathrm{MIT}=3.2$~meV/K (purple) and $\epsilon_\mathrm{MIT}=6.5$~meV/K (orange) for a fixed $T_\mathrm{MIT}=313$~K cross their respective counterparts (same color coding) at the transition temperature $T_c=338$~K, where both phases are equally likely. The fluctuating $R(T)$-traces simulated with a slow sweep rate of 1~mK/$\Delta t$ (lower panel, same color code) visualizes the increase in noise when the transition temperature of 338~K is approached. The relative noise levels versus voltage from a range of 34000 simulation points in the HRS (b) and LRS (c) for the different energy barriers from a).}
    \label{fig: barrier change}
\end{figure}
Thus, we focused the fitting of the simulation parameters solely on the phase fluctuations.

First, we analyze the influence of varying $\epsilon_\mathrm{IMT}$ and $\epsilon_\mathrm{MIT}$, while $T_\mathrm{IMT}$ and $T_\mathrm{MIT}$ remain constant.
The energy barriers for pairs of different $\epsilon_\mathrm{IMT}$ (solid lines) and $\epsilon_\mathrm{MIT}$ (dashed lines) with fixed $T_\mathrm{IMT}=357$~K and $T_\mathrm{MIT}=313$~K are shown in the upper panel of Figure \ref{fig: barrier change}~a. The vertical bar at $T_c=338$~K indicates the transition temperature where both phases become equally likely via the crossing of the energy barriers. The lower panel of Figure \ref{fig: barrier change}~a showcases the fluctuating $R(T)$-traces simulated with a slow sweep rate of 1~mK/$\Delta t$ which visualizes the increase in noise when the transition temperature is approached. Furthermore, it becomes apparent that large $\epsilon_\mathrm{IMT}$ and $\epsilon_\mathrm{MIT}$ (orange curve) confines the phase transition and fluctuation much closer to $T_c$ than small $\epsilon_\mathrm{IMT}$ and $\epsilon_\mathrm{MIT}$ (cyan curve). The voltage dependence of the relative noise in the HRS is shown in Figure \ref{fig: barrier change}~b. The resulting datasets reveal three changes when  $\epsilon_\mathrm{IMT}$ and $\epsilon_\mathrm{MIT}$ are varied, namely the amplitude of the noise, the voltage scaling and the switching voltage. First, let us discuss the amplitude variation. At a given temperature/voltage below $T_c$/$V_c$, a high difference of the energy barriers (orange circles) makes the HRS significantly more stable than the LRS and therefore the cells mostly remain in the HRS with rare transitions occurring, leading to a small noise signal. On the other hand, if the energy barriers are comparable in size (cyan circles), both states can be occupied and transitions between both states is common, leading to an increased noise signal. Second, the voltage scaling is more pronounced for higher $\epsilon_\mathrm{IMT}$ and $\epsilon_\mathrm{MIT}$, as an increase in voltage and Joule heat shifts the energy barrier stronger. As an example, we can assume barriers that are independent of the temperature/voltage and would therefore lead to the same noise level at all voltages. On the other hand, if the change of the energy barriers would be extremely steep, transition noise could only be observed at the proximity of $T_c$/$V_c$ where fluctuation between both phases occurs and would significantly drop when distancing from that point. Lastly, the decrease in the threshold voltage for lower $\epsilon_\mathrm{IMT}$ and $\epsilon_\mathrm{MIT}$ originates from the increase in current when more metallic cells are present and the resulting raise in Joule heat at same voltages compared to higher $\epsilon_\mathrm{IMT}$ and $\epsilon_\mathrm{MIT}$ where very few metallic cells are present.\\
Similarly to the HRS, the noise in the LRS close to the reset voltage, shown in Figure \ref{fig: barrier change}~c, also shows variations of the noise amplitude, the voltage scaling and the switching voltage. But the overall change when $\epsilon_\mathrm{IMT}$ and $\epsilon_\mathrm{MIT}$ is varied is consistently smaller compared to the HRS, as in the LRS there always exists a fluctuating interface close to $T_c$ (see schematics of Figure 5a,b in the main text), dominating the overall noise signal.
\begin{figure}[h!]
    \centering
    \includegraphics[width=\columnwidth]{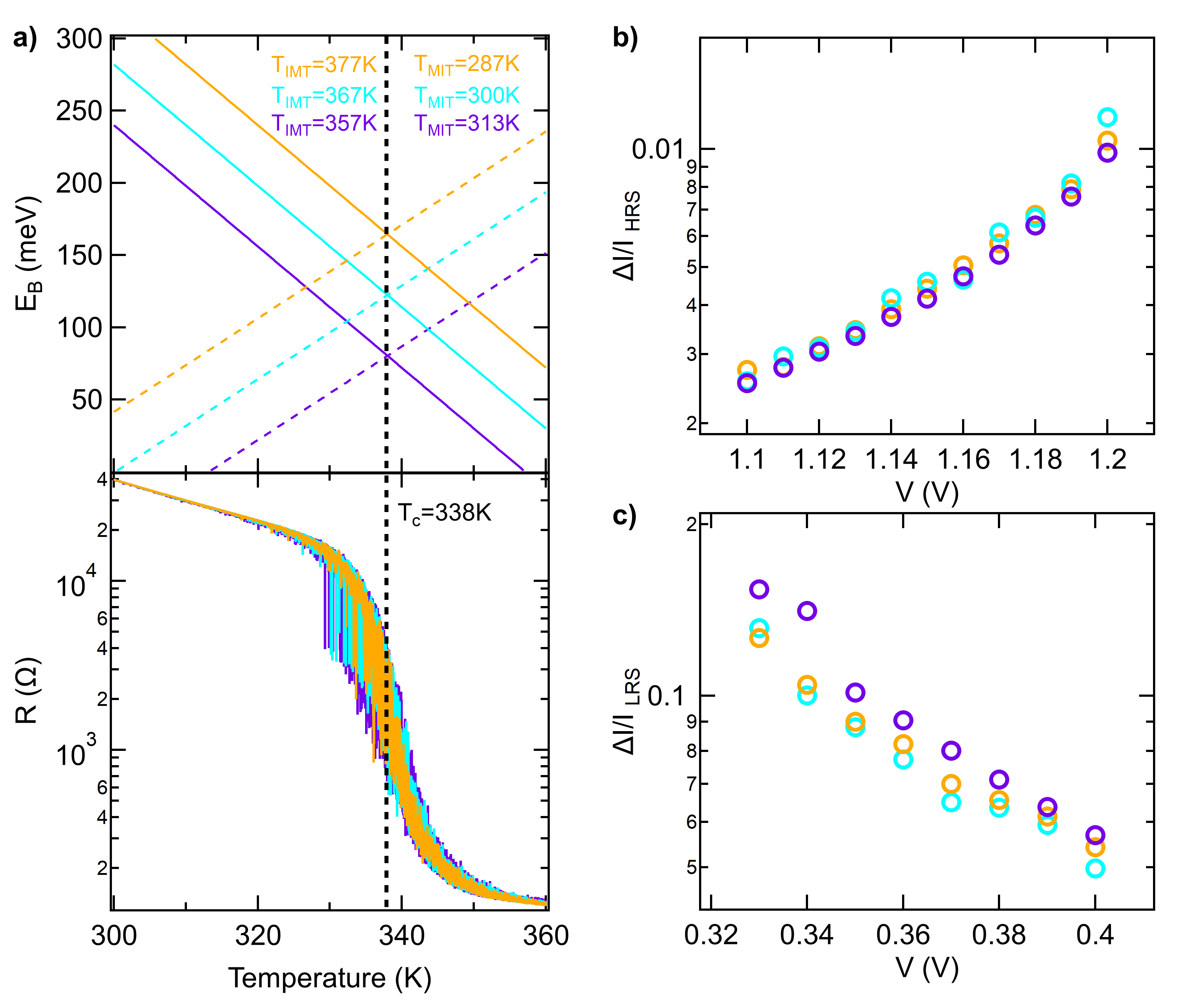}
    \caption{\textbf{Energy barrier offset comparisons} a) Comparison of different energy barriers (upper panel) for the insulator metal transition (solid lines) with $T_\mathrm{IMT}=357$~K (purple), $T_\mathrm{IMT}=367$~K (cyan) and $T_\mathrm{IMT}=377$~K (orange) for a fixed $\epsilon_\mathrm{IMT}=4.2$~meV/K. The reset barriers for the metal to insulator transition (dashed lines) with $T_\mathrm{MIT}=313$~K (purple), $T_\mathrm{MIT}=300$~K (cyan) and $T_\mathrm{MIT}=287$~K (orange) for a fixed $\epsilon_\mathrm{MIT}=3.2$~meV/K cross their respective counterparts (same color coding) at the transition temperature $T_c=338$~K, where both phases are equally likely. Variations of $T_\mathrm{IMT}$ and $T_\mathrm{MIT}$ are equivalent to an addition or subtraction of an offset to the barriers. The fluctuating $R(T)$-traces simulated with a slow sweep rate of 1~mK/$\Delta t$ (lower panel, same color code) visualizes the increase in noise when the transition temperature of 338~K is approached. The relative noise levels versus voltage from a range of 34000 simulation points in the HRS (b) and LRS (c) for the different energy barriers from a).}
    \label{fig: crosspoint change}
\end{figure}
After we analyzed the influence of $\epsilon_\mathrm{IMT}$ and $\epsilon_\mathrm{MIT}$ on the simulated noise signal, we then compared simulations with energy barriers, where $\epsilon_\mathrm{IMT}$ and $\epsilon_\mathrm{MIT}$ stay constant, while $T_\mathrm{IMT}$ and $T_\mathrm{MIT}$, where the barriers become zero, are varied. We can again arbitrarily choose a new $T_\mathrm{IMT}$, while $T_\mathrm{MIT}=\frac{\epsilon_\mathrm{IMT}(T_c-T_\mathrm{IMT})}{\epsilon_\mathrm{MIT}}+T_c$ is fixed by the other parameters. Such energy barriers are shown in the upper panel of Figure \ref{fig: crosspoint change}~a. It is important to realize that a shift of $T_\mathrm{IMT}$ and $T_\mathrm{MIT}$ is simply adding a constant offset to the energy barrier at any given temperature as \mbox{$E_\mathrm{IMT}=\epsilon_\mathrm{IMT}(T_\mathrm{IMT}-T)=\epsilon_\mathrm{IMT}(T'_\mathrm{IMT}\pm\Delta T_\mathrm{IMT}-T)=E'_\mathrm{IMT}\pm\mathrm{constant}$}. Similarly, the resulting switching probability is shifted by a multiplicative factor \mbox{$P_\mathrm{IMT} =\exp\left(\frac{-E_\mathrm{IMT}(T)}{\mathrm{k}_\mathrm{B}T}\right) =\exp\left(\frac{-E'_\mathrm{IMT}(T)\pm \mathrm{const.}}{\mathrm{k}_\mathrm{B}T}\right) =P'_\mathrm{IMT}\cdot\exp\left(\frac{\pm \mathrm{const.}}{\mathrm{k}_\mathrm{B}T}\right)$} which, close to the transition temperature $T_c$, can be described as $P_\mathrm{IMT}=\nu P'_\mathrm{IMT}$ with the attempt rate $\nu =\exp\left(\frac{\pm \mathrm{const.}}{\mathrm{k}_\mathrm{B}T_c}\right)$. As this shift in barrier energy and switching probability is equal for the IMT and MIT, the width of the transition and the fluctuating region of $R(T)$ remains the same as shown in the lower panel of Figure \ref{fig: crosspoint change}~a.

\begin{figure}[h!]
    \centering
    \includegraphics[width=0.8\columnwidth]{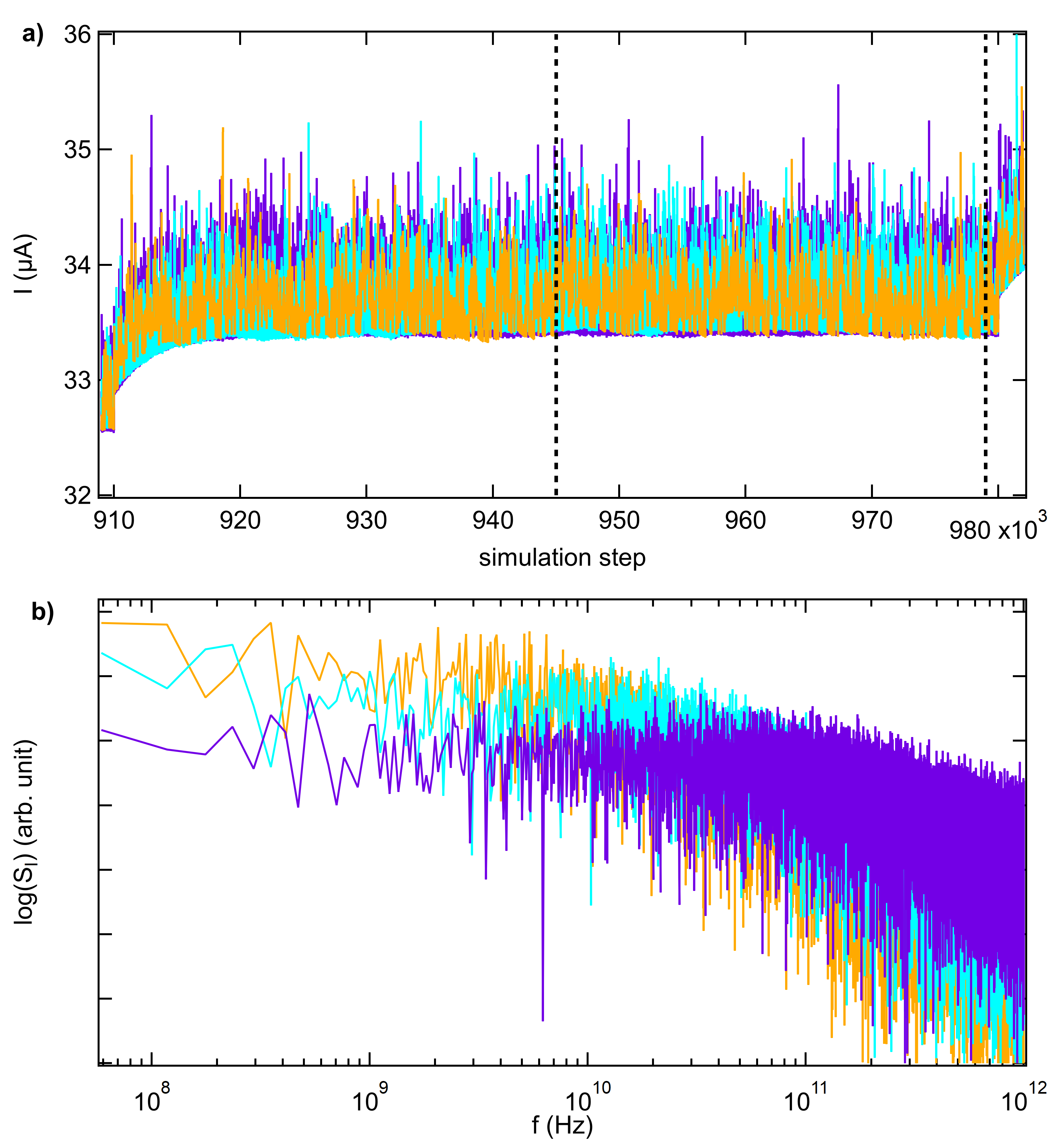}
    \caption{\textbf{Simulated current and PSDs} a) Simulated current in HRS for a 70000 points long $V_\mathrm{Drive}=1.18$~V for the three energy barriers from Figure \ref{fig: crosspoint change}. The vertical lines indicate the evaluation region for the PSDs in b). Increased energy barriers lead to decreased switching probabilities and decreases the cutoff frequency, shifting the noise power towards lower frequencies.}
    \label{fig: PSDs}
\end{figure}

The simulated voltage dependence of the noise, calculated from 34000 simulation steps at each voltage using the energy barriers of Figure \ref{fig: crosspoint change}~a, is shown in Figure \ref{fig: crosspoint change}~b (HRS) and c (LRS). As the signals are, within a margin of error, the same at all voltages, the shift of $T_\mathrm{IMT}$ and $T_\mathrm{MIT}$ has no influence on the noise signal. While this seems unintuitive at first as an increase (decrease) in barrier height decreases (increases) the transition rate, it simultaneously increases (decreases) the lifetime of the semi-stable phase showcased in the current fluctuation in Figure \ref{fig: PSDs}~a. Therefore, both effects cancel and the noise amplitude remains the same. The mathematical description of this cancellation can be explained by the power spectral density for the three barrier types shown in Figure \ref{fig: PSDs}~b. The three spectra were obtained from the simulated current timetraces at $1.18$~V for the HRS in the region between the 2 vertical lines of Figure \ref{fig: PSDs}~a, using the energy barriers from Figure \ref{fig: crosspoint change}~a. The simulated noise follows a Lorentzian spectra, which is dominated by a single time-constant $\tau$ and is described by $S_I=4A^2\frac{\tau}{1+\omega^2\tau^2}$, where $A$ is the noise amplitude and $\omega$ is the angular frequency. The total noise $\Delta I^2=\frac{1}{2\pi}\int_{\omega_1}^{\omega_2}S_I\mathrm{d}\omega=\frac{2}{\pi}A^2(\arctan(\omega_2\tau)-\arctan(\omega_1\tau))$, which for $\omega_2\gg\tau$ and $\omega_1\ll\tau$ becomes $\Delta I^2=A^2$, is independent of the time-constant $\tau$. Therefore, we can conclude that even though the timescale of the simulation $t=34000\cdot0.5~\mathrm{ps}=17$~ns and the frequency of the fluctuations $\frac{1}{\tau}$ are in the GHz range, the fundamental basics of the simulated fluctuations can be extended to any frequency range via an offset of the energy barriers $E_B$ and/or a shift in the attempt rate $\nu$ without changing the voltage dependence of $\frac{\Delta I}{I}$.

To summarize: First, we extracted the electrical and thermal parameters for the two-dimensional resistor network from experimental $R(T)$ and $R(V)$ datasets. Then, starting from a set of the 4 simulation parameters $\epsilon_\mathrm{IMT}$, $\epsilon_\mathrm{MIT}$, $T_\mathrm{IMT}$ and $T_\mathrm{MIT}$, we concluded that the transition temperatures, where the barrier becomes zero, only have an impact on the time-constant $\tau$ of the PSD but not the relative noise strength $\frac{\Delta I}{I}$ and kept them at the initial estimate of $T_\mathrm{IMT}=357$~K and $T_\mathrm{MIT}=313$~K, where the hysteresis in the experimental $R(T)$ vanishes. Then for a fixed $T_c=338$~K, choosing an arbitrary $\epsilon_\mathrm{IMT}$ fixates $\epsilon_\mathrm{MIT}$, which effectively brings the noise simulation down to one free fitting parameter. This single fitting parameter was then varied until the voltage dependence of the simulated and experimental noise matched. The experimental noise level $\frac{\Delta I}{I}$ of the device depends on the number of fluctuators as well as the investigated bandwidth. While, both could theoretically be reproduced in the simulation by decreasing the size of the simulated cells towards the unit cell of the VO$_2$ lattice and increasing the length (number of points) of the simulation to match the experimental time-scale, the necessary computational power would increase drastically. Therefore, we simply used a multiplicative scaling factor for the simulated noise level to match to the experimental level instead. In particular, this multiplicative factor does not change the voltage dependence of the simulated noise.

\section{Comparison of time- and frequency scale between the simulation and the experiment }

As mentioned in Section~\ref{sim_basics}, the $\Delta t$ time step in our simulation was set to $0.5$~ps in order to satisfy the thermal stability criterion, $\Delta t < C_i/G = \tau_{\mathrm{Thermal}}$. With this time step, reproducing the more than $200$~s long full-cycle noise measurement would require on the order of $10^{14}$ simulation steps, resulting in an unrealistic computational cost. Therefore, the actual timescale of the simulation is significantly shorter than the experimental timescale, corresponding to a simulation frequency bandwidth of $\sim 60~$MHz$~-~1~$THz, which differs significantly from the experimental bandwidth between $100~$Hz$~-~50~$kHz.

However, as previously demonstrated by the results shown in Figure~\ref{fig: crosspoint change} and Figure~\ref{fig: PSDs}, the amplitude of the relative noise is independent of the characteristic fluctuation time determined by the barrier height (or $T_{\mathrm{IMT}}$ and $ T_{\mathrm{MIT}}$), as long as the conditions $\tau \omega_1 \ll 1$ and $\tau \omega_2 \gg 1$ are fulfilled, i.e. the characteristic fluctuation frequency remains sufficiently far from the edges of the bandwidth. In the simulations, the $T_{\mathrm{IMT}}$ and $T_{\mathrm{MIT}}$ values (correspondingly the height of the energy barrier), were chosen based on the experimental $R(T)$-curve. The duration of the individual $V_{\mathrm{Drive}}$ plateaus was then selected such that, after the characteristic thermalization time of $\approx 5$~ns, sufficient time remained to achieve the required frequency resolution (i.e., $\omega_1 \ll 1/\tau$). Using different, but still realistic, energy barriers would result in characteristic $\tau$ values several orders of magnitude larger, corresponding to characteristic frequencies several orders of magnitude lower ($\approx 500~$meV to $700~$meV at $T_c$ for the 100~Hz to 50~kHz experimental frequency range), while leaving the qualitative behavior of the simulated curves unchanged. This would require long simulation time. On the other hand, the stability criterion defining $\tau_{\mathrm{Thermal}}$ scales with the cell size ($\tau_{\mathrm{Thermal}} \sim a^2$), therefore, increasing the cell size would allow larger $\tau_{\mathrm{Thermal}}$ values and thus larger simulation time steps. However, several orders of magnitude increase in the cell size would be required, which is not feasible for nanoscale devices. Consequently, the simulation timescale was kept in the picosecond range, while, based on the invariance discussed above, the obtained results remain valid also at experimentally relevant frequency scales.

It should also be noted that the shape of the simulated and experimental noise spectra differs: while the simulations yield Lorentzian spectra, the experimentally measured noise exhibits $1/f$-like behavior. This difference can be attributed to the simplified assumption in the simulations that the individual cells are identical and therefore characterized by similar fluctuation times $\tau$. In contrast, in the real system the local properties of the cells are not uniform, resulting in a distribution of the model parameters and consequently in a broader distribution of characteristic fluctuation times. The superposition of many fluctuators with different $\tau$ values leads to the $1/f$-type spectrum instead of a single Lorentzian contribution.

\section{Simplified parallel-electrode model for intrinsic noise scaling in the LRS}
To gain analytical insight into the resistance scaling of intrinsic noise in the LRS, where the active device volume (i.e., the metallic volume) evolves, we consider a simplified reference geometry consisting of a planar device with parallel electrodes separated by a distance $d$ and having a width $W$, which allows for a closed-form solution and an analytical estimate of the scaling exponent. The VO$_2$ layer between the electrodes is divided into cells, similarly to the 2D resistor network model, with dimensions $l_\mathrm{F} \times l_\mathrm{F} \times b$, where $l_F$ is the average length and width per fluctuator and $b$ is the thickness of the layer. Such a cell, containing a single fluctuator, can be characterized by the conductivity-parameters $(G_{\mathrm{F},LRS}, \Delta G_{\mathrm{F},LRS})$ and $(G_{\mathrm{F},HRS}, \Delta G_{\mathrm{F},HRS})$, depending on the state of the cell. The two types of cells occur in numbers $N_{LRS}$ and $N_{HRS}$, respectively, subject to the constraint $
N_{LRS} + N_{HRS} = N_{\mathrm{total}} = \frac{d\cdot W}{l_\mathrm{F}^2}$.
Taking into account that, in the LRS of the device, a given cell configuration can be described as rows of cells between the electrodes containing cells either in LRS or HRS, the total conductance of such a configuration can be obtained from Ohm’s law by connecting these rows in parallel:
\begin{equation}
G=\frac{W}{l_\mathrm{F}}\left[x\cdot G_{\mathrm{row},1} + (1-x)\cdot G_{\mathrm{row},2}\right],
\end{equation}
where $G_{\mathrm{row},1}= \frac{l_\mathrm{F}}{d} G_{\mathrm{F},LRS}$
and $G_{\mathrm{row},2}= \frac{l_\mathrm{F}}{d} G_{\mathrm{F},HRS}$
are the conductances of rows containing cells in LRS and HRS, respectively and $x=N_1/N_{\mathrm{total}}$ the ratio of cells in LRS to the total number of cells, which is equivalent to the width of the region in LRS.

In addition, the total conductance fluctuation can be calculated as
\begin{equation}
\Delta G^2 = \frac{W}{l_\mathrm{F}}\cdot\left[x\cdot \Delta G_{\mathrm{row},1}^2 + (1-x)\cdot \Delta G_{\mathrm{row},2}^2\right],
\end{equation}
where $\Delta G_{\mathrm{row},1}^2= \frac{l_{F}}{d}\cdot \Delta G_{\mathrm{F},LRS}^2$ ($\Delta G_{\mathrm{row},2}^2=\frac{l_{F}}{d}\cdot \Delta G_{\mathrm{F},HRS}^2$) is the conductance fluctuation of a row with cells in LRS (HRS).
Using the two equations above and the fact that in the linear I(V) regime $\frac{\Delta G}{G}=\frac{\Delta I}{I}$, the relative current fluctuation can be expressed as 
\begin{equation}
\frac{\Delta I}{I}=\sqrt{\frac{d}{W}}\cdot\frac{\Delta G_{\mathrm{F},LRS}}{G_{\mathrm{F},LRS}}\cdot\frac{\sqrt{x+(1-x)\cdot\frac{\Delta G_{\mathrm{F},HRS}}{\Delta G_{\mathrm{F},LRS}}}}{x+(1-x)\cdot\frac{G_{\mathrm{F},HRS}}{G_{\mathrm{F},LRS}}}.
\label{eq:DIpI_eff_A}
\end{equation}
Utilizing the experimentally observed resistance ratio and noise levels, $G_{\mathrm{F},LRS} >> G_{\mathrm{F},HRS}$ and $\Delta G_{\mathrm{F},LRS}^2 >> \Delta G_{\mathrm{F},HRS}^2$ can be considered and for $x \not\approx 0$, the formula above can be simplified to
\begin{equation}
\frac{\Delta I}{I}=\sqrt{\frac{d}{W}}\cdot\frac{\Delta G_{\mathrm{F},LRS}}{G_{\mathrm{F},LRS}}\cdot\frac{1}{\sqrt{x}}.
\label{eq:DIpI_eff_A2}
\end{equation}
By expressing $x$ in terms of the resistance as $x=\frac{d}{W\cdot G_{F,1}}\cdot G =\frac{d}{W\cdot G_{F,1}}\cdot R^{-1}$ and substituting it into the relative noise expression, the resistance dependence can be obtained as $\Delta I/I\sim R^{0.5}$.

The exponent of $0.5$ obtained from the parallel-electrode model can be interpreted in the context of the V-shaped electrode geometry by mapping the straight current paths between the electrodes onto curved trajectories as, illustrated in Fig.~5a of the paper, where the boundary of the metallic region can be well approximated by a circular arc. Under this approximation, and assuming a simplified geometry with smooth curved trajectories, the exponent remains $0.5$, consistent with the result of the parallel-electrode model.

In contrast, the two-dimensional resistor network simulations yield a slightly higher exponent of $0.59$. This deviation is attributed to the more accurate representation of the electrode geometry and to deviations of the metallic region from an ideal circular shape, both of which are captured in the numerical model but not in the simplified analytical description.

\section{Cycle-to-cycle variability of the threshold voltages}
\begin{figure}[h!]
    \centering
    \includegraphics[width=0.5625\columnwidth]{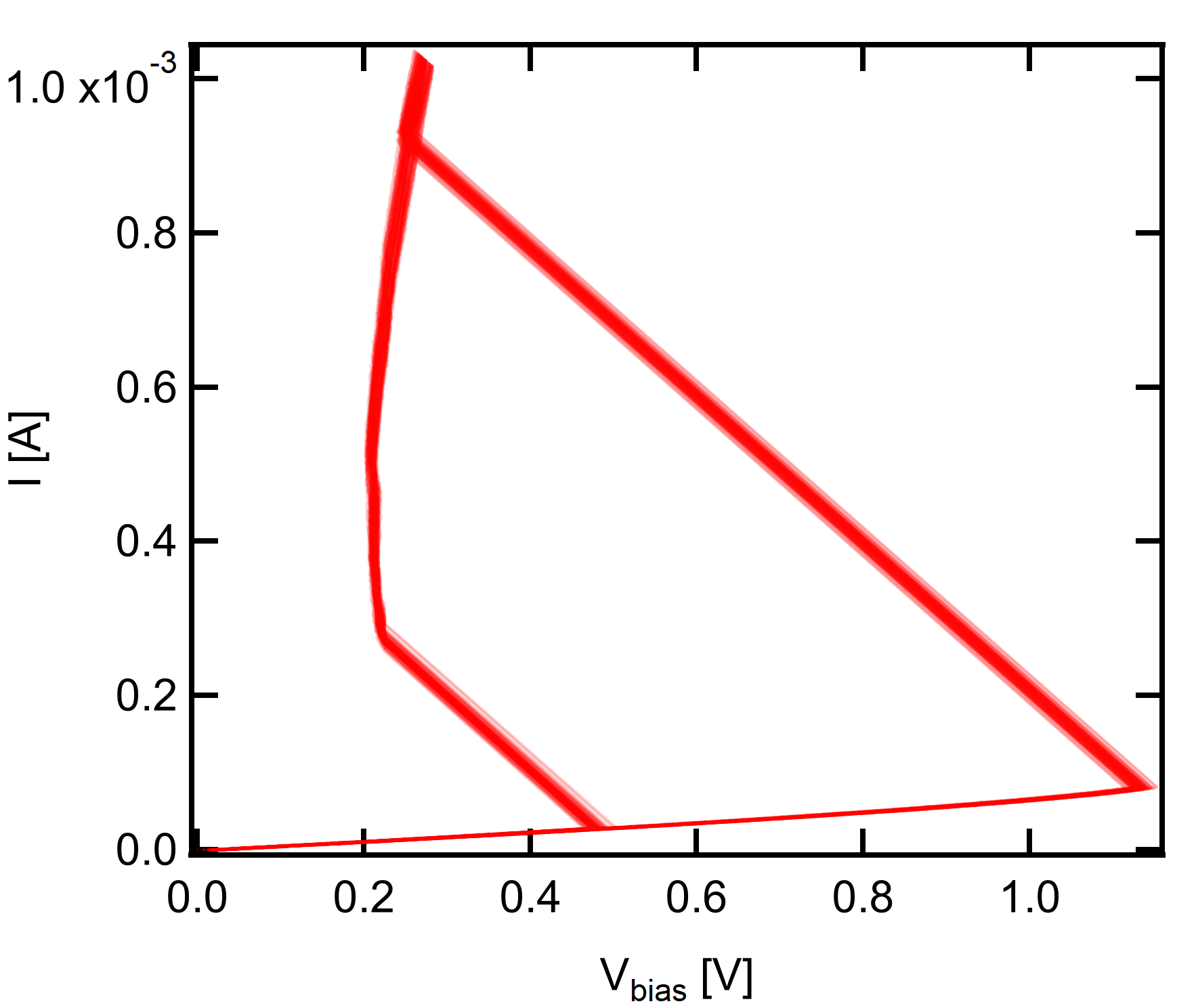}
    \caption{\textbf{I(V) characteristics over 80 consecutive switching cycles.} Semi-transparent red traces indicate individual cycles, with color intensity corresponding to the local occurrence probability. The resulting plot represents the distribution of the measured I(V) curves as a function of bias voltage. The measurements were performed using a $1050\,\Omega$ series resistance.}
    \label{fig:IVs}
\end{figure}

To investigate the variability of the switching threshold voltages ($V_\mathrm{Set}$ and $V_\mathrm{Reset}$), we carried out repeated I(V) measurements on our nanoscale VO$_2$ devices. Figure~\ref{fig:IVs} shows the current as a function of bias voltage in a representative measurement series of 80 consecutive cycles over the full voltage range. The broadening of the I(V) curves in the vicinity of the thresholds indicates cycle-to-cycle variations in the threshold voltages.

\begin{figure}[h!]
    \centering
    \includegraphics[width=0.8\columnwidth]{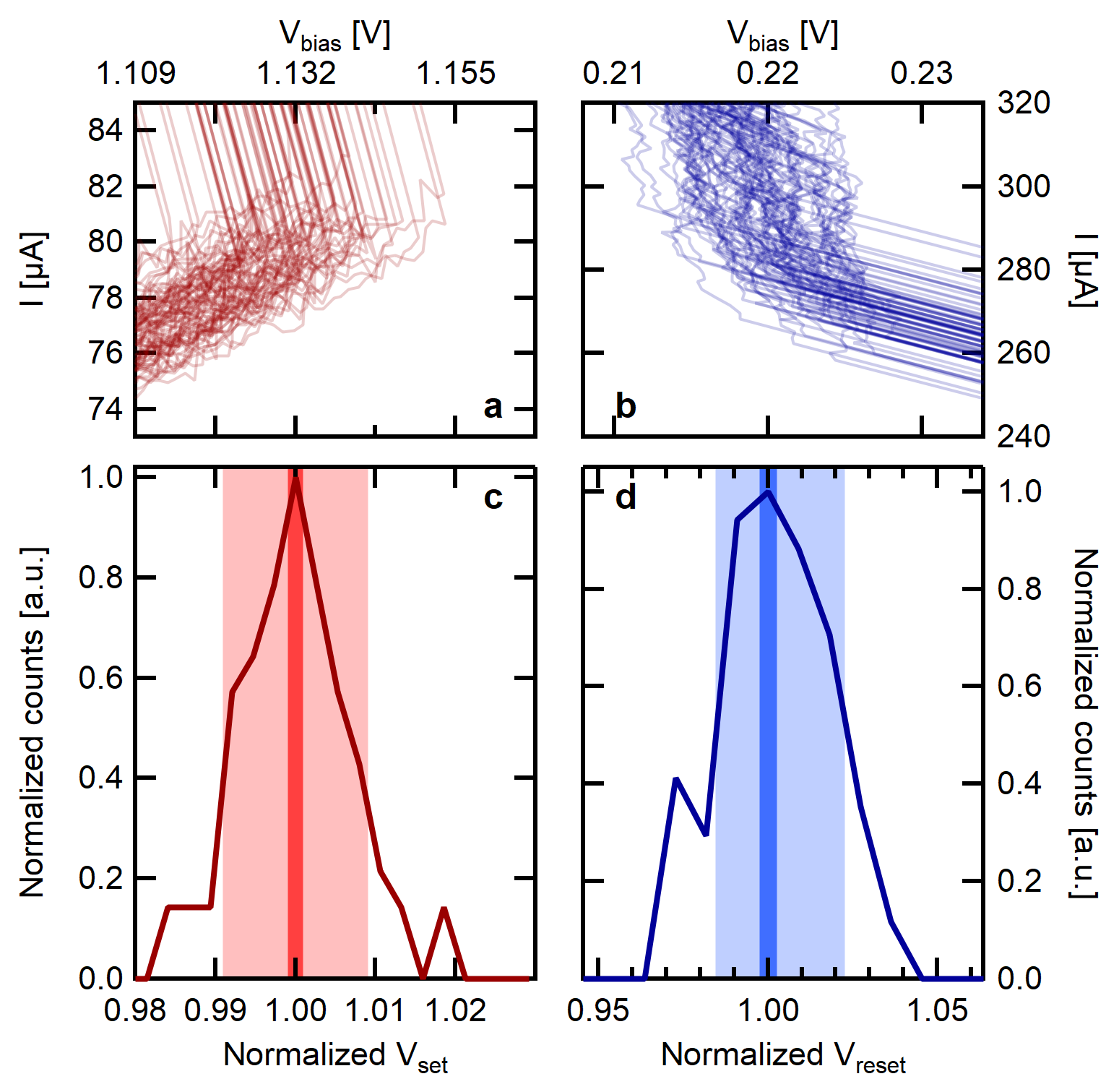}
    \caption{\textbf{Cycle-to-cycle statistics of switching threshold voltages.} Panels (a,c) show the set process, while panels (b,d) show the reset process. (a,b) Enlarged I(V) curves near the switching threshold shown as semi-transparent traces (dark red for set, blue for reset), as in Fig.~\ref{fig:IVs}. (c,d) Corresponding histograms of $V_\mathrm{Set}$ and $V_\mathrm{Reset}$ shown as dark red and blue curves, respectively, normalized by their most probable values (x-axis) and by their maximum counts (y-axis). The shaded and solid regions indicate the relative peak-to-peak current fluctuations and the corresponding relative standard deviations measured on the last voltage plateau before switching in the noise measurements.}
    \label{fig:thr_var}
\end{figure}

To highlight this variation, Figure~\ref{fig:thr_var}a,b shows the enlarged parts of the $I(V_\mathrm{Bias})$ curves near the $V_\mathrm{Set}$ ($V_\mathrm{Reset}$) voltages, while Figure~\ref{fig:thr_var}c,d shows the histograms of $V_\mathrm{Set}$ ($V_\mathrm{Reset}$) normalized by their most probable values. Based on the statistics, we find that the relative standard deviation of $V_\mathrm{Set}$ ($V_\mathrm{Reset}$) is approximately $0.7\%$ ($1.6\%$), and $90\%$ of the threshold data fall within a $2\%$ ($4.5\%$) relative range. These variations are consistent with the measured $1.8\%$ ($3.8\%$) relative peak-to-peak current fluctuations (see the shaded regions in Fig.~\ref{fig:thr_var}c,d) observed in the noise measurements on the last voltage plateau before the set (reset) process. This suggests that the observed transition noise, characterized by $0.2\%$ ($0.5\%$) relative standard deviation and $\Delta I / I \approx 8 \cdot 10^{-4}$ ($1 \cdot 10^{-3}$) relative current fluctuations, is consistent in magnitude with the threshold voltage variability, indicating that it can impose a lower bound on the achievable threshold variance.

\newpage

\bibliography{references}
\bibliographystyle{achemso.bst}